\documentclass[11pt]{article}
\usepackage{amssymb,amsmath,amsfonts,epsfig,color,latexsym}
\def\appendix{{\newpage\section*{Appendices}}\let\appendix\section%
        {\setcounter{section}{0}
        \gdef\thesection{\Alph{section}}}\section}

\newcommand{\pa}{\partial}
\newcommand{\be}{\begin{equation}}
\newcommand{\ee}{\end{equation}}
\newcommand{\bea}{\begin{eqnarray}}
\newcommand{\eea}{\end{eqnarray}}

\newcommand{\T}{\mbox{Tr}}
\newcommand{\bD}{\bar{D}}

\newcommand{\bP}{\bar\Phi}

\newcommand{\cN}{{\cal N}}
\newcommand{\cK}{{\cal K}}
\newcommand{\cO}{{\cal O}}

\renewcommand{\P}{\Phi}
\renewcommand{\a}{\alpha}

\newcommand{\da}{\dot\alpha}
\renewcommand{\b}{\beta}

\newcommand{\q}{\theta}
\newcommand{\bq}{\bar\q}

\newcommand{\ep}{\epsilon}

\newcommand{\bt}[1]{{\bar t}}

\begin{document}

\thispagestyle{empty}
\null \hfill AEI-2004-067\\
\null\vskip-12pt \hfill ROM2F-04/09 \\
\null\vskip-12pt \hfill  LAPTH-1061/04 \\
\vskip0.2truecm
\begin{center}
\vskip 0.2truecm {\Large\bf
\Large{A three-loop test of the dilatation operator in $\cN=4$ SYM }
}\\
\vskip 1truecm
{\bf B. Eden$^{*,**}$, C. Jarczak$^{***}$ and E. Sokatchev$^{***}$ \\
}

\vskip 0.4truecm
$^{*}$ {\it Max-Planck-Institut f\"ur Gravitationsphysik,
Albert-Einstein-Institut, \\
Am M\"uhlenberg 1, D-14476 Golm, Germany}\\
\vskip .2truecm $^{**}$ {\it Dipartimento di Fisica, Universit\`a di Roma ``Tor Vergata"\\
I.N.F.N. - Sezione di Roma ``Tor Vergata" \\
Via della Ricerca Scientifica, 00133 Roma, Italy}\\
\vskip .2truecm $^{***}$ {\it Laboratoire d'Annecy-le-Vieux de
Physique Th\'{e}orique  LAPTH,\\
B.P. 110,  F-74941 Annecy-le-Vieux, France\footnote{UMR 5108 associ{\'e}e {\`a}
 l'Universit{\'e} de Savoie} } \\
\end{center}

\vskip 1truecm \Large
\centerline{\bf Abstract} \normalsize We compute the three-loop anomalous dimension of the BMN operators with charges $J=0$ (the Konishi multiplet) and $J=1$ in the $\cN=4$ super-Yang-Mills theory. We employ a method which effectively reduces the calculation to two loops. Instead of using the superconformal primary states, we consider the ratio of the two-point functions of suitable descendants of the corresponding multiplets. \\
Our results unambiguously select the form of the $\cN=4$ SYM dilatation operator which is compatible with BMN scaling. Thus, we provide evidence for BMN scaling at three loops.

\newpage
\setcounter{page}{1}\setcounter{footnote}{0}

\section{Introduction}

During the last couple of years, following the important work of BMN \cite{Berenstein:2002jq}, spectacular progress has been made in the understanding of the integrable structure underlying $\cN=4$ supersymmetric Yang-Mills (SYM) theory \cite{Minahan:2002ve,Beisert:2003tq,Beisert:2003yb}, on the one hand, and its correspondence to string theory \cite{Frolov:2002av,Beisert:2003xu,Arutyunov:2004xy}, on the other. An important part of this program is the development of efficient tools for calculating anomalous dimensions of gauge invariant composite operators and then comparing to string theory predictions.

One such tool is the so-called ``dilatation operator" of $\cN=4$ SYM \cite{Beisert:2002ff,Beisert:2003tq}. In this approach, without doing complicated higher-loop perturbative calculations, one is able to predict the anomalous dimensions of large classes of operators. The form of the dilatation operator can be determined by combining various symmetry arguments with some field theory input. In \cite{Beisert:2003tq} the notion of one-loop integrability was extended to higher loops. Based on this and on the additional assumption of BMN scaling, the planar dilatation operator of the $SU(2)$ sector of $\cN=4$ SYM was elaborated up to three loops. Later on, a comprehensive treatment of the dilatation operator for the $SU(2|3)$ sector was given in \cite{Beisert:2003ys}.  It was shown that the two-loop dilatation operator is determined up to one free constant, which can be fixed from the known value of the two-loop anomalous dimension of the Konishi operator.  At three loops, after taking into account the known quantum symmetries, a two-parameter freedom remains. The higher-loop integrability conjectured in \cite{Beisert:2003tq} was shown to follow from conformal supersymmetry and the dynamics of the theory, keeping the above mentioned parameters arbitrary.

In  \cite{Beisert:2003ys} Beisert used the extra assumption of BMN scaling to fix the remaining freedom. Alternatively, ``experimental data" in the form of the three-loop anomalous dimensions of any two BMN operators, obtained by a direct perturbative calculation, would unambiguously determine the three-loop dilatation operator. This would also provide an indirect check on the hypothesis of BMN scaling at this level of perturbation theory.\footnote{At two loops BMN scaling has already been confirmed in \cite{Gross:2002su}, but it seems rather difficult to extend this result to three loops. A general argument to all orders was proposed in \cite{Santambrogio:2002sb}, but it is based on a number of assumptions which, although plausible, seem hard to justify. Also, an indication of BMN scaling at three loops through the matrix model has appeared in \cite{Klose:2003qc}.}

Thus, the necessity of such a three-loop perturbative  ``experiment" appears
quite clear. Rather recently, a whole series of three-loop anomalous
dimensions of operators of twist two were computed in
\cite{Vogt:2004mw,Kotikov:2004er}, using sophisticated QCD methods. The
simplest among these operators is the Konishi scalar, which is also the lowest
state in the BMN series. The value of its anomalous dimension obtained in
\cite{Kotikov:2004er} is in perfect agreement with the prediction of
\cite{Beisert:2003tq,Beisert:2003ys} under the assumption of BMN
scaling. However, it is not clear if the method of
\cite{Vogt:2004mw,Kotikov:2004er} generalizes to operators of higher twist,
such as the scalar BMN operators with non-vanishing charge.\footnote{Here we
  mean the twist and the charge  of the primary state of a superconformal
  multiplet. Thus, the Konishi multiplet starts with a scalar of dimension two
  and charge zero, so it has twist two and $J=0$; the scalar BMN operators
  with two impurities of
  charge $J$ have dimension and twist $J+2$, etc. This differs from the
  classification of \cite{Berenstein:2002jq}, where the dimension and the
  charge are shifted by two units by considering a certain descendant of the multiplet. } So, obtaining the second piece of ``experimental data" needed for the above test is still an open problem.

In this paper we carry out a three-loop calculation of the anomalous dimension
of the first two operators in the BMN series (with two impurities), the Konishi one (i.e., twist two or $J=0$) and the twist three (or $J=1$) one. In the Konishi case we reproduce the result of \cite{Kotikov:2004er} (unlike Ref. \cite{Kotikov:2004er}, we use standard field theory techniques). The result for the $J=1$ case is new. Most remarkably, the two values are exactly those predicted by the dilatation operator whose form is compatible with BMN scaling!

Direct three-loop calculations of anomalous dimensions are prohibitively complicated. Here we exploit a method based on an idea of Anselmi \cite{Anselmi:1998ms} and further refined and applied at two loops in \cite{Eden:2003sj}. It allows us to effectively reduce the perturbative order to two loops.\footnote{Loop counting for composite operators is not the same as counting powers of the coupling $g$. Thus, the ``tree-level" two-point function of an $n$--linear composite operators contains $n-1$ loops, but no interactions, so it is $O(g^0)$. In this article, as is customary in the AdS/CFT literature, ``n loops" means $O(g^{2n})$. However, the actual loop integrals to be evaluated may be of higher order, see Section \ref{334}.} In this section we give a brief description of the basic ideas.

In conformal field theory operators are labeled by their
spin and dilatation weight. The two-point function of a scalar
operator has the functional form
\begin{equation}
\langle \, \cO(x) \, \cO^\dagger(0) \, \rangle \, = \,
\frac{c(g^2)}{(x^2)^{\Delta(g^2)}} \, , \label{form1}
\end{equation}
with
\begin{equation}
\Delta(g^2) \, = \, \Delta_0 + \gamma(g^2)\, .
\end{equation}
Here $\Delta_0$ is the canonical (or tree-level) scaling dimension and $\gamma(g^2) = \sum^\infty_{n=1}\gamma_n \, g^{2n}$ is the ``anomalous" part of the dilatation weight. The latter is due to logarithmic divergences of the two-point function which, after proper renormalization of the quantum operator $\cO$, are supposed to sum up into the power behavior indicated in eq. (\ref{form1}).

 The idea of Anselmi's trick is most easily understood on the example of a ``current" $k_\mu(x)$ possessing anomalous dimension $\Delta = 3 + \gamma(g^2)$. Conformal invariance fixes its two-point function up to normalization:
\begin{equation}\label{00}
  \langle \, k_\mu(x) \, k_\nu^\dagger(0) \, \rangle \, = \,
c(g^2)\, \frac{x^2\eta_{\mu\nu} - 2 x_\mu x_\nu}{(x^2)^{4 + \gamma(g^2)}}
\end{equation}
In the free field theory, where $g=0$ and $\gamma(g^2)=0$, this current is conserved, so
\begin{equation}\label{01}
  \langle \, \partial^\mu k_\mu(x) \, \partial^\nu k_\nu^\dagger(0) \, \rangle_{g=0} \, = \, 0\,,
\end{equation}
up to contact terms which we do not consider here. In the presence of (quantum) interaction $\gamma(g^2)\neq 0$ and we find
\begin{equation}\label{001}
  \langle \, \partial^\mu k_\mu(x) \ \partial^\nu k_\nu^\dagger(0) \, \rangle_{g\neq 0} \, = \, \, +4 \, c(g^2)\, \frac{\gamma(\gamma+2)}{(x^2)^{4 + \gamma(g^2)}}\,.
\end{equation}
Now, consider the ratio
\begin{equation}\label{0011}
\frac{\langle \, \partial^\mu k_\mu(x) \ \partial^\nu k_\nu^\dagger(0) \, \rangle}{\langle \, k^\rho(x) \, k_\rho^\dagger(0) \, \rangle}
   \, = \, \, +2 \, \frac{\gamma(\gamma+2)}{x^2}\,.
\end{equation}
Notice that the right-hand side of this equation is $O(g^2)$, since $\gamma(g^2) = \gamma_1 g^2 + O(g^4)$, and so must be the left-hand side. This is indeed true, given the fact that $\partial^\mu k_\mu = O(g)$ as a consequence of the field equations (i.e., conservation in the absence of interaction). In practical terms this means that we have gained an order in $g^2$: If we wished to compute, e.g., $\gamma_1$ starting from eq. (\ref{00}), we would have to evaluate {\it one-loop} divergent integrals involved in the two-point function of the current itself. However, using the ratio (\ref{0011}) instead, it is sufficient to compute the {\it tree-level} two-point functions of the current and of its divergence. It is precisely this trick that Anselmi  \cite{Anselmi:1998ms} applied to the current in the Konishi multiplet in order to obtain its one-loop anomalous dimension, without doing any loop integrals!

The generalization to higher orders in perturbation theory is complicated by the quantum anomaly phenomenon. By this we mean that the non-conservation of the current may not be correctly explained by the classical field equations alone, but it might be modified by an extra, purely quantum term. This is exactly what happens to  the Konishi current, as we explain below.

In the $\cN=4$ SYM theory the Konishi current and its divergence are  members of one ``long" (or unprotected) superconformal multiplet. The lowest component (or superconformal primary state) of the Konishi superfield $\cK_{\mathbf{1}}(x,\q,\bq)$ is an $SU(4)$ singlet scalar  of dimension 2. This superfield contains $15+1$ Konishi currents, $\cK_{\mathbf{1}} = \ldots + \q_A\sigma^\mu\bq^B\, (k_\mu)^A{}_B(x)+ \ldots\,$, where $A,B=1,2,3,4$ are $SU(4)$ indices. The divergence in (\ref{001}) is contained in the double spinor derivative of the superfield $\bD_A \cdot \bD_B \ \cK_{\mathbf 1}$ (here $\bD_A \cdot \bD_B = \bD_B \cdot \bD_A = -1/4 \, \bD_{\dot\alpha A} \bD^{\dot\alpha}_B$). Taking such derivatives and subsequently using the superfield equations of motion is equivalent to making on-shell supersymmetry transformations. In the free theory $\bD_A \cdot \bD_B \ \cK_{\mathbf 1} =0$, which implies conservation of the type (\ref{01}), $\partial^\mu (k_\mu)^A{}_B=0$. In the interacting theory this equation is modified, $\bD_A \cdot \bD_B \ \cK_{\mathbf 1} = g\,(\cK_{\mathbf{10}})_{AB}$. Its right-hand side gives rise to a new scalar superfield of dimension 3 in the \textbf{10} of $SU(4)$, which contains the divergence  $\partial^\mu k_\mu(x)$. This new scalar is a superconformal descendant of the Konishi multiplet \cite{Intriligator:1999ff}, \cite{Bianchi:2001cm}. Notice the presence of the gauge coupling $g$ in the definition of $\cK_{\mathbf{10}}$, indicating that this descendant does not occur in the free theory. Now, in the $\cN=4$ theory we can go a step further, applying two more spinor derivatives (i.e.,  on-shell supersymmetry transformations) to produce a scalar of dimension 4 in the \textbf{84}, $g^2\,(\cK_{\mathbf{84}})^{CD}{}_{AB} =  D^{C}\cdot D^{ D}\ \bD_{A} \cdot \bD_{B}
 \, \cK_{\mathbf 1}- \mbox{(traces)}$. This time we have a factor of $g^2$ in the definition of $\cK_{\mathbf{84}}$, showing that we have used the field equations twice.

The supersymmetric version of eq. (\ref{0011}) is the ratio of two-point functions
\begin{equation}\label{0R'}
\frac{\langle \, \cK_{\mathbf{10}} \, \cK_{\mathbf{10}}^\dagger \,
\rangle}{\langle \, \cK_{\mathbf{1}} \, \cK_{\mathbf{1}}^\dagger \, \rangle}\,.
\end{equation}
In principle, it contains information about the anomalous dimension common to all members of the Konishi multiplet (which include $\cK_{\mathbf{1}}$ and $\cK_{\mathbf{10}}$) However, the procedure is not safe beyond the lowest perturbative level. The reason is that the naive definition of the descendant $\cK_{\mathbf{10}}$ obtained by applying the {\it classical} field equations (or on-shell supersymmetry) to the {\it quantum} (renormalized) operator $\cK_{\mathbf{1}}$ is not correct. What this naive procedure gives is the classical ``anomaly" $\cK_{\mathbf{10}} = B$, where $B$ is the superpotential of the $\cN=4$ theory. The procedure misses, however, another term $gF$, where $F$ is a composite operator with the same quantum numbers (dimension and $SU(4)$ representation), but this time made out of fermions. This is the so-called quantum Konishi anomaly \cite{Clark:1978jx}. So, the correct form of the descendant (schematically) is $\cK_{\mathbf{10}} = B + g F$. Even knowing this, the correct evaluation of the ratio (\ref{0R'}) is still not obvious. The problem is to properly fix the normalization of the quantum corrected descendant $\cK_{\mathbf{10}}$. Recall that when going from (\ref{00}) to  (\ref{001}), we tacitly assumed that the normalization $c(g^2)$ was the same; this is far from obvious when the divergence of the current is affected by a quantum anomaly. So, it might seem that the ``Anselmi trick" is of little use beyond one loop.

The way out was proposed and successfully used at order $O(g^4)$ in \cite{Eden:2003sj}. The idea is to evaluate another ratio,
\begin{equation}\label{0R}
\frac{\langle \, \cK_{\mathbf{84}} \, \cK_{\mathbf{84}}^\dagger \,
\rangle}{\langle \, \cK_{\mathbf{10}} \, \cK_{\mathbf{10}}^\dagger \, \rangle}\,.
\end{equation}
Why is this better? We can argue that the second step in the chain of descendants above, $\cK_{\mathbf{10}} \ \to \ \cK_{\mathbf{84}}$, is anomaly free. We identify the origin of the Konishi ``anomaly" with the operator mixing taking place at the level of the descendant $\cK_{\mathbf{10}}$. As we just mentioned, at this level there are two possible composite operators with the same quantum numbers, $B$ and $F$, which can and do mix between themselves. The correct definition of   $\cK_{\mathbf{10}}$ is achieved by resolving this mixing problem (see Section \ref{boost} and \cite{us} for a detailed discussion). On the other hand, at the level of the descendant $\cK_{\mathbf{84}}$ there is only one candidate, a single-trace operator of dimension 4 (see again Section \ref{boost}).\footnote{To be more precise, there is also a double-trace version of it which could possibly mix. However, one can show (see \cite{Bianchi:1999ge} for a one-loop and \cite{Bianchi:2000hn} and Appendix \ref{AppB} to the present paper for a two-loop proof) that this mixing only affects another operator, a member of a protected ``quarter-BPS" multiplet; the Konishi descendant $\cK_{\mathbf{84}}$ stays pure.} So, we can conclude that the absence of mixing at the level of $\cK_{\mathbf{84}}$ implies the absence of the anomaly. In this case, we assume that the step $\cK_{\mathbf{10}} \ \to \ \cK_{\mathbf{84}}$ can be done naively, using the classical superfield equations of motion or on-shell supersymmetry. In particular, this means that $\cK_{\mathbf{84}}$ ``inherits" its normalization from its ancestor $\cK_{\mathbf{10}}$, so that it drops in the ratio (\ref{0R}).

Apart from the central issue of anomalies, considering the ratio (\ref{0R}) rather than  (\ref{0R'}) has another, technical but very important advantage. We plan to do our Feynman graph calculations in the $\cN=1$ superspace formalism. There $\cK_{\mathbf{1}}$ is a real (non-chiral) superfield, whereas for both $\cK_{\mathbf{10}}$ and  $\cK_{\mathbf{84}}$ we can find an $SU(4)$ projection which is chiral.\footnote{One should not confuse ``chiral operators" with ``chiral primary operators" (CPO). The latter are known to be protected from quantum correction. The chiral projections of the Konishi descendants we are considering are not protected, precisely because they are not superconformal primary. For this reason we prefer the more adequate term ``BPS operators" instead of the popular term CPO.} This significantly reduces  the number of Feynman graphs involved. Further, evaluating the ratio instead of the two-point functions themselves results in cancellation of many Feynman integrals. Finally, the set of graphs for the two-point function of the chiral projection has a large overlap with those for a certain half-BPS operator. The graphs for the latter are known to sum up to zero (``protectedness"), which allows us to eliminate large sets of graphs without calculating them (see also \cite{Arutyunov:2002jg,Bianchi:2003eg}).

This method applies not only to the Konishi multiplet, but also to any multiplet which lies on the unitarity bound of the continuous series of $\cN=4$ superconformal representations \cite{Dobrev:1985qv}, \cite{Ferrara:2000eb}. Their common property is that they satisfy a generalized ``conservation" condition $D^2 \cO=0$ in the free case. In the presence of interaction some of these multiplets still satisfy this condition, and so they have protected conformal weight (these are the so-called ``semishort" multiplets \cite{SemiShort,Bianchi:2001cm}). However, the generic multiplet of this type is not conserved in the interacting theory, so it is not protected and acquires an anomalous dimension. It is precisely this ``non-conservation" that we wish to exploit according to the scenario described above. In principle, the entire series of BMN multiplets falls into this class,\footnote{We emphasize the point that only the Konishi multiplet contains an anomalous vector current. The higher-twist BMN multiplets do not contain such currents, but nevertheless satisfy an anomalous ``semishortness" condition in superspace. Thus, focusing on the analogy between the anomaly of the Konishi current and the standard axial anomaly is somewhat misleading, in our opinion.} but in practice the quantum calculation are further complicated by operator mixing already at the level of the superconformal primaries. Only the first two members of the BMN family, the $J=0$ (or twist two) multiplet (which is the Konishi multiplet) and the $J=1$ (or twist three), are realized in terms of pure composite operators (without mixing). This is why we limit the scope of our three-loop calculations to these two cases. As explained earlier, our results are sufficient to unambiguously confirm the form of the $\cN=4$ dilatation operator  proposed in \cite{Beisert:2003tq,Beisert:2003ys}.

The paper is organized as follows. In Section \ref{boost} we recall the general properties of the Konishi multiplet and its descendants. We explain in detail how superspace differentiation reduces the loop order. We also give the realization of the various objects as gauge invariant composite operators both in $\cN=4$ and in $\cN=1$ superspace. We proceed to determine $\gamma_2$ and $\gamma_3$ from a graph calculation: in Section \ref{s3} this is done for the Konishi case. In Section \ref{L6} we adapt the calculation, with relatively minor changes, to the case of the BMN multiplet with $J=1$. In the Conclusions we summarize the results obtained and discuss their implications. The Appendix contains our conventions and superspace Feynman rules, discusses the absence of mixing for
the level two descendants, and comments on the use of the $\overline{MS}$
scheme.

\section{Reducing the loop order by superspace differentiation}\label{boost}

\subsection{The Konishi operator and its descendants}\label{ss1}

The Konishi operator is the simplest example of a composite gauge invariant operator in the $\cN=4$ super Yang-Mills theory. The basic object in this theory is the super-field-strength $W_{AB}(x^\mu,\q^\a_A,\bq^{\da\, A})=-W_{BA}$ where $A,B=1,2,3,4$ are indices of the fundamental representation of the R symmetry group $SU(4)$. The lowest component of this superfield describes six real scalars due to the reality condition
\begin{equation}\label{1}
\overline{(W_{AB})} \, = \, W^{AB} \, = \, - \frac{1}{2} \,
\epsilon^{ABCD} \, W_{CD} \, .
\end{equation}
In addition, $W_{AB}$ is subject to the on-shell constraints
\begin{equation}\label{2}
  \nabla^{\{C}_\a W_{A\}B} = 0\,, \qquad \bar\nabla_{\da\, (C} W_{A)B}=0\,,
\end{equation}
where ${}^{\{C}{}_{A\}}$ denotes the traceless part and $(CA)$ means symmetrization with unit weight.
The gauge-covariant superspace derivatives  $\nabla,\ \bar\nabla$ satisfy the algebra
\begin{eqnarray}
  \{\nabla^A_\a , \nabla^B_\b   \} &=&  - 4 \, g \, \ep_{\a\b} W^{AB}\,,\nonumber\\
  \{\nabla^A_\a , \bar\nabla_{\da\, B}   \} &=& - 2 \, i \, \delta^A_B\, \sigma^\mu_{\a\da}\nabla_\mu   \label{4}
\end{eqnarray}
(we have explicitly written down the gauge coupling $g$). A useful corollary of eqs. (\ref{2}) and (\ref{4}) is
\begin{equation}\label{5}
  \nabla^{(A} \cdot \nabla^{B)}\, W^{CD} = - g [W^{(AC},W^{B)D}]\,,
\end{equation}
where $\nabla^A \cdot \nabla^B \, = \, -1/4 \, \nabla^{A \a} \nabla_{\a}^B$.

The (on-shell) Konishi operator is defined by
\begin{equation}\label{6}
  \cK_{\mathbf 1} = \frac{1}{4} \T \left(W^{AB}W_{AB}  \right)\,.
\end{equation}
Its lowest component (at $\q=\bq=0$) is a scalar of naive dimension 2. This scalar is an $SU(4)$ singlet, as indicated by the subscript ${\mathbf 1}$. The ${\cal N}=4$ supersymmetric generalization of the two-point function (\ref{form1}) for $\cK_{\mathbf 1}$ is (see \cite{Park:1997bq} for the $\cN=1$ case)
\begin{equation}
\langle \, \cK_{\mathbf 1}(x_1,\q_1, \bq_1) \, \cK_{\mathbf 1}(x_2,\q_2,\bq_2) \, \rangle \, = \,
\frac{c(g^2)}{(\hat x_{L1R2}^2 \, \hat
x_{L2R1}^2)^{\Delta(g^2)/2}} \label{form2}
\end{equation}
with the chiral/antichiral supersymmetric invariant coordinate
\begin{equation}\label{7}
\hat x_{L1R2}^\mu \, =  \, x_{1}^\mu - x_{2}^\mu + i \theta_{1A} \sigma^\mu \bar \theta^A_1 + i \theta_{2A} \sigma^\mu \bar \theta^A_2 - 2 i \theta_{1A} \sigma^\mu \bar \theta^A_2 \,.
\end{equation}
Clearly, (\ref{form2}) reduces to (\ref{form1})
when $\theta, \bar \theta = 0$.

Applying spinor derivatives to the superfield $\cK_{\mathbf 1}$, we can produce all of its components
 (or ``descendants"). In this paper we exploit the following two descendants \cite{Intriligator:1999ff}, \cite{Bianchi:2001cm}:
\begin{eqnarray}
 g\, (\cK_{\mathbf{10}})_{AB} & = & \bD_A \cdot \bD_B \ \cK_{\mathbf 1} |_{\theta, \, \bar
 \theta \, = \, 0} \, , \label{8} \\
 g^2\, (\cK_{\mathbf{84}})^{CD}{}_{AB} & = & D^{\{C}\cdot D^{\{ D}\ \bD_{A\}} \cdot \bD_{B\}}
 \ \cK_{\mathbf 1} |_{\theta, \, \bar \theta \, = \, 0} \,.  \label{9}
\end{eqnarray}
In eq. (\ref{9}) we are taking the traceless part only, so the order of the derivatives $D$ and $\bD$ does not matter. It is clear that the descendants (\ref{8}) and (\ref{9}) are scalars of naive dimension 3 and 4, in the ${\mathbf{10}}$ (Dynkin labels $[200]$) and ${\mathbf{84}}$ ($[202]$) of $SU(4)$, respectively. Notice that we write down the gauge coupling $g$ explicitly, in order to remind ourselves that these descendants do not occur (or rather decouple) in the free theory.

Further, for reasons which will become clear below, we find it very convenient to work with the highest weight projections of the above $SU(4)$ representations:\footnote{The $SU(4)$ representations can be read off the highest weight by applying the following rule: Count the occurrences of the lower indices $1,2,3$ and fill in the boxes of a Young tableau (the first row with 1, etc). When a lower index 4 occurs, it can be removed by simultaneously dropping a full set $1234$; an upper index 4 is replaced by a lower set $123$.  Thus, the Young tableau corresponding to ${\cal K}_{\mathbf{10}}$ in (\ref{des}) has labels $(200)$ which translates into Dynkin labels $[200]$, i.e., the ${\mathbf{10}}$. For ${\cal K}_{\mathbf{84}}$ we have ${}^{44}_{11}\ \rightarrow\ 1111\; 22\; 33$, which is the Young tableau $(422)$ or Dynkin labels $[202]$, i.e., the ${\mathbf{84}}$.}
\begin{eqnarray}
g\, {\cal K}_{\mathbf{10}} & \equiv &  g\, ({\cal K}_{\mathbf{10}})_{11} = \bar D_1 \cdot \bar D_1 \, {\cal K}_{\mathbf{1}} |_{\theta, \, \bar
\theta \, = \, 0} \, , \label{des} \\
g^2\, {\cal K}_{\mathbf{84}} & \equiv &  g\, ({\cal K}_{\mathbf{84}})^{44}{}_{11} = D^4 \cdot D^4
\, \bar D_1 \cdot \bar D_1 \, {\cal K}_{\mathbf{1}} |_{\theta, \, \bar \theta \, = \, 0}
\, . \nonumber
\end{eqnarray}
Their two-point functions can be found from (\ref{form2}) by
applying the same spinor derivatives as in the definition of
the operators:
\begin{eqnarray}
g^2\, \langle \, \cK_{\mathbf{10}}(x_1) \, \cK_{\mathbf{10}}^\dagger(x_2) \,
\rangle & = & \bar D_1 \cdot \bar D_1 |_1 \, D^1 \cdot D^1 |_2 \, \langle \, {\cal K}_{\mathbf{1}}(1) \, \cK_{\mathbf 1}(2) \, \rangle|_{\theta, \, \bar \theta
\, = \, 0} \label{mess}  \, , \\
g^4\, \langle \, \cK_{\mathbf{84}}(x_1) \, {\cal K}_{\mathbf{84}}^\dagger(x_2) \, \rangle & = & D^4 \cdot D^4 \, \bar D_1 \cdot \bar D_1 \, |_1 \, \bar D_4
\cdot \bar D_4 \, D^1 \cdot D^1 \, |_2 \, \langle \, {\cal K}_{\mathbf{1}}(1) \, \cK_{\mathbf 1}(2) \, \rangle|_{\theta, \, \bar \theta
\, = \, 0} \nonumber \, .
\end{eqnarray}
Due to chirality, $\bar D_1 \cdot \bar D_1 |_1 \, D^1 \cdot D^1 |_2$ only
acts on $\hat x_{L2R1}$. The superspace differentiation results into $-\square$
on the corresponding factor in the two-point function. The second step is
analogous: $D^4 \cdot D^4 |_1 \, \bar D_4 \cdot \bar D_4 |_2$ only sees
$\hat x_{L1R2}$, producing another $-\square$.  Hence
\begin{eqnarray}
g^2\, \langle \, \cK_{\mathbf{10}}(1) \, \cK_{\mathbf{10}}^\dagger(2) \,
\rangle & = &  - \Delta (\Delta-2) \,
\frac{c(g^2)}{(x_{12}^2)^{(\Delta(g^2)+1)}} \, , \label{expo} \\
g^4\, \langle \, \cK_{\mathbf{84}}(1) \, \cK_{\mathbf{84}}^\dagger(2) \, \rangle
& = & \Delta^2 (\Delta-2)^2 \,
\frac{c(g^2)}{(x_{12}^2)^{(\Delta(g^2)+2)}} \, . \nonumber
\end{eqnarray}

Let us consider the ratios of the two-point functions in (\ref{form2}) (at $\q=\bq=0$) and in (\ref{expo}):
\begin{eqnarray}
  \frac{\langle \, \cK_{\mathbf{10}}  \, \cK_{\mathbf{10}}^\dagger  \, \rangle} {\langle \, \cK_{\mathbf{1}} \, \cK_{\mathbf{1}}^\dagger
\, \rangle} &=& - \frac{\Delta (\Delta-2)}{g^2\, x^2_{12}} \label{ratio1} \, , \\
  \frac{\langle \, \cK_{\mathbf{84}}  \, \cK_{\mathbf{84}}^\dagger  \, \rangle} {\langle \, \cK_{\mathbf{10}} \, \cK_{\mathbf{10}}^\dagger
\, \rangle} &=& - \frac{\Delta (\Delta-2)}{g^2\, x^2_{12}}  \label{ratio2} \, .
\end{eqnarray}
Apart from the trivial space-time factor, the right-hand sides contain information about the anomalous dimension,
\begin{equation}\label{anodi}
  \frac{\Delta(\Delta-2)}{g^2} = 2\gamma_1 + (\gamma_1^2 + 2\gamma_2)g^2 + 2(\gamma_1\gamma_2+\gamma_3)g^4 + O(g^6)\,.
\end{equation}We clearly see the ``loop reduction" effect of the superspace differentiation: It is sufficient to compute the left-hand side of either eq. (\ref{ratio1}) or eq. (\ref{ratio2}) to order $O(g^{2(n-1)})$ (``$n-1$ loop") in order to determine the ``$n$--loop" anomalous dimension $\gamma_n$ via a simple algebraic equation.

\subsection{The descendants as composite operators }\label{s22}

In the free field theory ($g=0$) the descendants of $\cK_{\mathbf 1}$ considered above do not occur. Instead, the free Konishi superfield satisfies the conservation (or ``semishortness") condition
\begin{equation}\label{10}
   D^A \cdot D^B \ \cK_{\mathbf 1}\vert_{g=0} = \bD_A \cdot \bD_B \ \cK_{\mathbf 1}\vert_{g=0} = 0\,,
\end{equation}
as easily follows from the on-shell constraints (\ref{2}), (\ref{4}). In particular, this implies that the ${\mathbf{15}}+{\mathbf{1}}$ vector components of dimension 3 in $\cK_{\mathbf{1}}$ are conserved vectors (the so-called ``Konishi currents"). However, this conservation is destroyed when the interaction is switched on. Indeed, applying once again the on-shell constraints (\ref{2}), (\ref{4}), but this time with $g\neq 0$, after some simple algebra we find
\begin{equation}\label{11}
  \bD_A \cdot \bD_B \ \cK_{\mathbf 1}\vert_{g \neq 0} = \frac{g}{2} \; \T\left([W_{AC}, W_{BD}] W^{CD} \right) \equiv \, - 3 g \, B_{AB}\,.
\end{equation}
This is the so-called ``classical Konishi anomaly" reflecting the fact that the ``conservation" (\ref{10}) does not correspond to any symmetry of the interacting theory.

In the quantum theory eq. (\ref{11}) is further modified by the  fermionic term
\begin{equation}\label{12}
  F_{AB} = \frac{1}{9} \, \T\left(\nabla^{\a\, C} W_{CA}\; \nabla^{D}_\a W_{DB} \right)\,,
\end{equation}
so that
\begin{equation}\label{13}
  \bD_A \cdot \bD_B \ \cK_{\mathbf 1}\vert_{\rm quantum} = -3 g \, \bigl(
B_{AB} + g f_{10} \; F_{AB} \bigr)  + O(g^3) \equiv g(\cK_{\mathbf{10}})_{AB}\,.
\end{equation}
The coefficient $f_{10}= \frac{N}{32 \pi^2}$ of the $F$ term (the so-called ``quantum Konishi anomaly") has been calculated at one loop in \cite{Clark:1978jx}, but the operator relation (\ref{13}) receives higher-order corrections indicated by the $O(g^3)$ terms.\footnote{In the literature there are claims that the Konishi anomaly is one-loop exact, like the Adler-Bell-Jackiw anomaly. We believe that this statement is strongly context- and interpretation-dependent. In \cite{us} this issue is discussed in the context of operator mixing and it is shown that the anomaly receives corrections.} So, the correct definition of the quantum operator--{\it descendant} of the Konishi operator is
\begin{equation}\label{13'}
  (\cK_{\mathbf{10}})_{AB} = -3 \, \bigl( Z_B\,  B_{AB} + Z_F \,  F_{AB} \bigr)
\, ,
\end{equation}
where $Z_B, Z_F$ are renormalization factors specified in Section \ref{s31} (see also \cite{us} for a more detailed discussion).
We stress that eq. (\ref{8}) simply defines a {\it component} of the superfield $\cK_{\mathbf 1}$. The difference between a true descendant and a superfield component is that the former is obtained through the (quantum corrected) field equations and the latter through straightforward superspace differentiation.

The next level descendant $\cK_{\mathbf{84}}$ is obtained by further differentiation (see (\ref{9})):
\begin{equation}\label{14}
  g\, (\cK_{\mathbf{84}})^{CD}{}_{AB} = D^{\{C}\cdot D^{\{ D}\; (\cK_{\mathbf{10}})_{A\}B\}}
\end{equation}
and use of the on-shell constraints. This is most easily done in terms of the highest weight projections as defined in (\ref{des}). The corresponding projections of the bosonic and fermionic composite operators $B$ and $F$ are
\begin{eqnarray}
  B &\equiv& B_{11} = \T \left([W_{12},W_{13}]W_{14}  \right)\nonumber \, , \\
  F &\equiv& F_{11} = \frac{1}{9} \, \T\left(\nabla^{\a\, C} W_{C1}\; \nabla^{D}_\a W_{D1} \right)\,. \label{14'}
\end{eqnarray}
Applying the derivatives $D^4 \cdot D^4$ and using the constraints, we find
\begin{equation}\label{susyBF}
D^4 \cdot D^4 \, B \, = \, g \, Y \, , \qquad
D^4 \cdot D^4 \, F \, = \, 4 g^2 \, Y \, ,
\end{equation}
where
\begin{equation}\label{15}
  Y = \T\left([W_{12},W_{13}] [W_{12},W_{13}] \right)
\end{equation}
is a scalar operator of dimension 4 corresponding to the highest weight of the ${\mathbf{84}}$. We see that both ingredients $B$ and $F$ of $\cK_{\mathbf{10}}$ give rise to the same operator $Y$, so we are led to the conclusion
\begin{equation}\label{16}
g\, \cK_{\mathbf{84}} \, = \, D^4 \cdot D^4 \, \cK_{\mathbf{10}} \, = \, g \,
\left( Z_B \, + \, 4 g \, Z_F \right) \, Y \, \equiv \, g \,
Z_{\mathbf{84}} \, Y\,.
\end{equation}

Here we ought to make an important comment. We obtained the {\it quantum} operator $\cK_{\mathbf{84}}$ from the {\it quantum} operator $\cK_{\mathbf{10}}$ through the {\it classical} field equations. However, at the preceding step $\cK_{\mathbf{1}}\ \rightarrow \ \cK_{\mathbf{10}}$ we saw that the same procedure only produced the $B$ term (\ref{11}), it completely missed the $F$ term in (\ref{13}). Can we be sure that something similar does not happen at the step $\cK_{\mathbf{10}}\ \rightarrow \ \cK_{\mathbf{84}}$? Our argument is as follows: the reason why the quantum $\cK_{\mathbf{10}}$ is different from the classical one is operator mixing. Indeed, there exist two operators with the same quantum numbers (scalars of dimension 3 in the $\mathbf{84}$), our $B$ and $F$. In the quantum theory they can mix, therefore the correct form of $\cK_{\mathbf{10}}$ should be established by resolving this mixing problem (see Section \ref{s3} and \cite{us}). At first sight, at the level of $\cK_{\mathbf{84}}$ we seem to be in a similar situation. Besides the single-trace operator $Y$ (\ref{15}), there is also a double-trace version of it,
\begin{equation}\label{17}
  Y' = \T \left(W_{12}W_{12}\right)\; \T \left(W_{13}W_{13}\right) - \T \left(W_{12}W_{13}\right)\; \T \left(W_{12}W_{13}\right)
\end{equation}
(the antisymmetrization is needed to achieve the right Young tableau structure of the \textbf{84}).
The mixing of $Y$ and $Y'$ has been studied at one loop in \cite{Bianchi:1999ge,Bianchi:2002rw,Ryzhov:2001bp} and the conclusion is that the Konishi descendant $\cK_{\mathbf{84}}$ is indeed identified with $Y$, while a particular mixture of $Y$ and $Y'$ gives rise to a different, protected (quarter-BPS) operator which is orthogonal to $\cK_{\mathbf{84}}$. For our purposes here we need to make sure that the same picture persists at the next, $O(g^4)$ level, and this is done in Appendix \ref{AppB} (see also \cite{Bianchi:2000hn}). We view this as a sufficient reason to assume that the step $\cK_{\mathbf{10}}\ \rightarrow \ \cK_{\mathbf{84}}$ can be done naively, so that (\ref{16}) is the exact form of the operator $\cK_{\mathbf{84}}$.

\subsection{The stress-tensor multiplet}\label{s23}

We remark that the two equations (\ref{susyBF}) imply the existence of a particular combination of $B$ and $F$ which has a vanishing descendant in the ${\mathbf{84}}$:
\begin{equation}\label{17'}
\cO_{\mathbf{10}} \, = \, -3(F \, - 4 g \, B) \quad \rightarrow \quad D^4 \cdot D^4 \, \cO_{\mathbf{10}}=0 \,.
\end{equation}
In fact, the operator $\cO_{\mathbf{10}}$ is itself a descendant of the protected (half-BPS) operator $\cO_{\mathbf{20'}}$, the so-called ``stress-tensor multiplet" of $\cN=4$ SYM:
\begin{equation}\label{18}
  (\cO_{\mathbf{20'}})_{ABCD} = \T\left(W_{AB}W_{CD} + W_{AD}W_{CB}  \right)
\end{equation}
Note that this combination has vanishing traces of the type $(\cO_{\mathbf{20'}})_{ABCD}\; \ep^{BCDE} = 0$, so it indeed corresponds to the $\mathbf{20'}$ ([020]) in the decomposition $\mathbf{6}\times \mathbf{6} = \mathbf{20'} + \mathbf{15} + \mathbf{1}$. The important property of the operator (\ref{18}) is that it satisfies the half-BPS constraints
\begin{eqnarray}
  && D^{\{E}_\a(\cO_{\mathbf{20'}})_{A\}BCD} = D^{\{E}_\a(\cO_{\mathbf{20'}})_{AB\}CD} = 0 \nonumber \, , \\
  && \bD_{\da\, (E}(\cO_{\mathbf{20'}})_{A)BCD} = \bD_{\da\, (E}(\cO_{\mathbf{20'}})_{AB)CD} = 0 \label{19}  \, ,
\end{eqnarray}
which follow from the on-shell constraints on $W$. This is true even in the presence of interactions, and as a consequence, the conformal dimension of $\cO_{\mathbf{20'}}$ is ``protected" from quantum corrections. Moreover, it has been shown \cite{Eden:1999gh,Penati:1999ba,Penati:2000zv} that the two-point function of this operator is not renormalized (up to contact terms),
\begin{equation}\label{19'}
  \langle \, \cO_{\mathbf{20'}} \, \cO_{\mathbf{20'}}^\dagger\, \rangle \, = \, \langle \, \cO_{\mathbf{20'}} \, \cO_{\mathbf{20'}}^\dagger\, \rangle_{g=0}\,.
\end{equation}

The descendant $\cO_{\mathbf{10}}$ is obtained in the following way:
\begin{equation}\label{20}
  (\cO_{\mathbf{10}})_{AB} = D^C\cdot D^D (\cO_{\mathbf{20'}})_{ACBD} =  \, -3
\, \bigl(F_{AB} - 4 g \, B_{AB} \bigr) \,,
\end{equation}
and eq. (\ref{17'}) corresponds to the highest weight projection $(\cO_{\mathbf{10}})_{11}$. Another consequence of the half-BPS conditions (\ref{19}) is that this multiplet is ``short", i.e., it does not have the full span of descendants for a generic $\cN=4$ multiplet. This explains why the step indicated in (\ref{17'}) does not produce a new descendant.

The two-point function of the descendant $\cO_{\mathbf{10}}$, like the one of the primary (\ref{19'}), remains unchanged by quantum corrections,
\begin{equation}\label{21}
  \langle \, \cO_{\mathbf{10}} \, \cO_{\mathbf{10}}^\dagger\, \rangle \, = \, \langle \, \cO_{\mathbf{10}} \, \cO_{\mathbf{10}}^\dagger\, \rangle_{g=0}\,.
\end{equation}
For our graph calculations in Section \ref{s3} it will be essential that not only the two-point function of $\cO_{\mathbf{10}}$, but also its explicit form (\ref{20}) as a mixture of the bare operators $F$ and $B$ is not renormalized. This is not obvious, since it is not easy to rule out the possibility of finite corrections in an operator mixture, even if it has no anomalous dimension. In \cite{us} a perturbative test of this conjecture has been carried out.

The two operators $\cO_{\mathbf{10}}$ and $\cK_{\mathbf{10}}$ belong to different superconformal multiplets, the former to a ``short" one with protected (canonical) dimension, the latter to a ``long" one with anomalous dimension. Therefore they must be orthogonal to each other,
\begin{equation}\label{22}
  \langle \, \cO_{\mathbf{10}} \, \cK_{\mathbf{10}}^\dagger\, \rangle \, = \, 0\,.
\end{equation}
The two properties (\ref{21})
and (\ref{22}) are all that we need in order to resolve the mixing of the operators $B$ and $F$. We just have to prepare the right mixtures which satisfy these two conditions (see Section \ref{s3} and \cite{us}).

\subsection{The BMN operators}

The Konishi multiplet is the first member of the family of BMN operators (here we consider the case of two ``impurities" only). Let us give a brief description of it.

The stress-tensor multiplet can be generalized to a family of half-BPS operators \cite{HoweWest} of dimension $J$ in the representation $[0J0]$ of $SU(4)$. Their highest weights can be written in the symbolic form
\begin{equation}\label{BPS}
   \cO_{[0J0]} = (W_{12})^J
\end{equation}
(the color trace (single or multiple) is understood). They satisfy half-BPS shortening conditions similar to (\ref{19}). After projecting the $SU(4)$ indices these conditions read (for details see \cite{Ferrara:2000eb})
\begin{equation}\label{BPScon}
   D^4\;\cO_{[0J0]} = D^3\;\cO_{[0J0]} = \bD_1\;\cO_{[0J0]} = \bD_2\;\cO_{[0J0]} = 0\,.
\end{equation}
Their meaning is that the superfield $\cO_{[0J0]}$ depends on half of the Grassmann variables (hence the name ``half-BPS"; such superfields are also called ``Grassmann analytic''),
\begin{equation}\label{BPStheta}
   \cO_{[0J0]}(\q_a, \bq^{a'})\,, \quad a =1,2\,, \ a'= 3,4\,.
\end{equation}

The BMN family of operators with two ``impurities" is obtained by ``gluing" together $\cO_{[0J0]}$ with the Konishi operator $\cK_{\mathbf{1}}$, thus constructing a new scalar operator of dimension $J+2$ in the $[0J0]$. Symbolically,\footnote{The full picture is more complicated, since the two constituents of the Konishi operator can be positioned in different ways under the common color trace. Also, an operator with the same quantum numbers can be realized using fermions of the type $\nabla^A W_{AB}$. Here we work in the simplest cases $J=0,1$ where these issues are irrelevant.}
\begin{equation}\label{BMN}
   \cK_{[0J0]} = \cO_{[0J0]}\times \cK_{\mathbf{1}}\,.
\end{equation}
where $\times$ means merging the two color traces. Now, combining the free theory semishortness conditions (\ref{10}) with the
BPS conditions (\ref{BPScon}), we see that free BMN operators satisfy the intersection of the two, i.e. the semishortness conditions (only the highest weight is shown)
\begin{equation}\label{BMNsemi}
   D^4 \cdot D^4 \, \cK_{[0J0]} = \bD_1 \cdot \bD_1 \, \cK_{[0J0]} = 0\,,
\end{equation}
which corresponds to a superconformal representation at the unitarity bound of the continuous series. In the interacting case, just like for Konishi, these conditions give rise to descendants (recall (\ref{des}))
\begin{eqnarray}
g\, {\cal K}_{[2J0]} & = & \bar D_1 \cdot \bD_1 \, {\cal K}_{[0J0]} |_{\theta, \, \bar \theta \, = \, 0} \, , \label{desJ} \\
g^2\, {\cal K}_{[2J2]} & = & D^4 \cdot D^4 \, \bar D_1 \cdot \bar D_1 \,
{\cal K}_{[0J0]} |_{\theta, \, \bar \theta \, = \, 0}
\, . \nonumber
\end{eqnarray}

The two-point functions of the operators ${\cal K}_{[0J0]}$ are determined by $\cN=4$ conformal supersymmetry as follows. First, consider the two-point function of the half-BPS operator (\ref{BPS}):
\begin{equation}\label{2pBPS}
   \langle \cO_{[0J0]}(1)\; \cO_{[0J0]}^\dagger (2)\rangle = \frac{C}{(\hat{x}_{1\bar{2}}^2)^J}  \,.
\end{equation}
Here $C$ is a normalization constant (or a constant $SU(4)$ tensor, if not restricted to the highest weight only). The supersymmetric invariant difference (c.f. (\ref{7}) in the chiral/antichiral case)
\begin{equation}
\hat{x}_{1\bar{2}} = x_1-x_2 + i\bigl(\q_{1a}\bq^{a}_1 - \q_{1a'}\bq^{a'}_1 +
\q_{2a}\bq^{a}_2 - \q_{2a'}\bq^{a'}_2   -2\q_{1a}\bq^{a}_2 +2 \q_{2a'}\bq^{a'}_1 \bigr)
 \label{diffe}
\end{equation}
is the unique combination compatible with the BPS conditions (\ref{BPScon}). What remains to do is to multiply the structure (\ref{form2}) for $\cK_{\mathbf{1}}$ with that for $\cO_{[0J0]}$ (\ref{2pBPS}), according to (\ref{BMN}):
\begin{equation}\label{2pBMN}
   \langle \cK_{[0J0]}(1)\; \cK_{[0J0]}^\dagger (2)\rangle = \frac{c(g^2)}{(\hat x_{L1R2}^2 \, \hat
x_{L2R1}^2)^{\Delta(g^2)/2}\, (\hat{x}_{1\bar{2}}^2)^J}  \,,
\end{equation}
where $\Delta = 2 + \gamma(g^2)$. Thus, the ``Konishi factor" carries the anomalous dimension, while the ``half-BPS factor" is there to adjust the canonical dimension $\Delta_0 = 2 + J$.

It is now rather obvious that repeating the same differentiation as in (\ref{mess}), we obtain results for the descendants ${\cal K}_{[2J0]}$ and ${\cal K}_{[2J2]}$ similar to (\ref{expo}), since the BPS factor is annihilated by these derivatives.

In this paper we are only interested in $J=0$ and $J=1$. We restrict to
$SU(N)$ gauge group so that there is in both cases only one gauge invariant
composite operator with the right quantum numbers, the Konishi operator
$\cK_{\mathbf{1}}$
(\ref{6}) and
\begin{equation}\label{K6}
   (\cK_{\mathbf{6}})_{AB} = \frac{1}{2} \,\T \left(W_{AB} W^{CD} W_{CD}  \right)  \, ,
\end{equation}
respectively. We thus avoid the complications of operator mixing on the
level of the primaries.

\subsection{Realization in terms of $\cN=1$ superfields}\label{s24}

The quantum calculations we have in mind are most easily carried out in $\cN=1$ superspace. To this end we need to translate the $\cN=4$ expressions obtained above into $\cN=1$ notation.

The on-shell ${\cal N} = 4$ gauge multiplet contains the fields $
\phi_{[AB]}, \, \psi_{A \, \alpha}, \, \bar \psi^A_{\dot \alpha},
\, A_\mu $. The six scalar fields $\phi_{AB} = W_{AB} |_{\q,\bq \, =\, 0}$
obey the reality condition (\ref{1}). The spinor components are defined by
$\psi_{A \, \alpha} \, = \, 1/3 \, \nabla^B_\a \, W_{BA} |_{\q,\bq \,= \,0}$.

The transition from $\cN=4$ to $\cN=1$ is accompanied by a breakdown of R symmetry, $SU(4)\ \rightarrow SU(3)\times U(1)$. Accordingly, we can organize the  three complex scalars and three of the four spinors into on-shell ${\cal N} = 1$ chiral matter multiplets:
\begin{equation}
\Phi^I \, = \, \phi_{1, \,I+1} + \theta^\alpha \, \psi_{\alpha,
\, I+1}\,, \qquad I=1,2,3 \, ,
\end{equation}
where we have identified $\q=\theta_1$. The remaining fields $A, \, \lambda,
\, \bar \lambda $ (where we have set  $\lambda \, = \, \psi_1$) constitute the on-shell vector multiplet of $\cN=1$ SYM. In the full ${\cal N}=1$ formulation these are completed to off-shell chiral
superfields $\Phi^I(x_L, \q)$ and a real Yang-Mills gauge superfield $V(x,\q,\bq)$ by a set of
auxiliary and pure gauge fields. Replacing the gauge field $A_\mu$ by its field strength $F_{\mu\nu}$, we can form another chiral superfield, the $\cN=1$ on-shell super-field strength:
\begin{equation}\label{22'}
  W_\a = \lambda_\a + \q^\b (\sigma^{\mu\nu})_{\a\b} F_{\mu\nu}\,.
\end{equation}

\subsubsection{The operator $\cK_{\mathbf{1}}$}

Let us now give the $\cN=1$ expressions of the various gauge invariant
operators from the preceding subsections. The Konishi superfield is
\begin{equation}\label{23}
  \cK_{\mathbf{1}} = \T \left(e^{gV} \bar \Phi_I e^{-gV} \Phi^I  \right)\,.
\end{equation}
The role of the exponential factors is to make the composite operator gauge
invariant. The highest weight projection $B$ (\ref{14'}) of the operator in
the $\mathbf{10}$ is expressed in terms of matter superfields of the same
chirality:
\begin{equation}\label{24}
B \, = \, \T([\Phi^1, \Phi^2] \Phi^3)\,,
\end{equation}
while the operator $F$ is given by the chiral $\cN=1$ field strength:
\begin{equation}\label{25}
  F \, = \, \T (W^\a W_\a)\,.
\end{equation}
The highest weight projection $Y$ (\ref{15}) of the operator in the $\mathbf{84}$ is again chiral:
\begin{equation}\label{26}
Y \, = \, \T([\Phi^1, \Phi^2][\Phi^1, \Phi^2])\,.
\end{equation}

We now clearly see the advantage of working with the highest weight projections. In the $\cN=1$ language they all correspond to chiral operators. In particular, there is no need for gauge factors like in the Konishi operator (\ref{23}). The quantum calculations with such operators are significantly easier than with projections of mixed chirality, e.g., $\T([\Phi^1, \Phi^2] \bar\Phi_3)$.

A quick way for constructing descendants is to use the on-shell ${\cal N}=4$ supersymmetry transformations of the component fields (we only  list the relevant part):
\begin{eqnarray}
\delta \phi_{AB} & = & - \eta_{[A}^\alpha \psi_{B] \, \alpha} +
\frac{1}{2} \, \epsilon_{ABCD} \, \bar \eta^{[C}_{\dot \alpha}
\bar \psi^{D] \, \dot \alpha} \, , \nonumber \\ \delta \psi_{A
\, \alpha} & = & - \eta_A^\beta \sigma^{\mu\nu}_{(\alpha \, \beta)} F_{\mu\nu}
- g \, \eta_{B\, \alpha} \, [\phi_{AC}, \phi^{BC}] +
i \, \bar \eta^{B \, \dot \beta} D_{\alpha \, \dot \beta} \, \phi_{AB} \, .
\label{dPhi}
\end{eqnarray}
(here $[A,B]$ exceptionally means antisymmetrization with weight one). In the preceding sections we used the superspace derivatives, e.g., $D^4 \cdot D^4$, which is equivalent to making two supersymmetry transformations with parameter $\eta_4$. With the help of the rules
\begin{eqnarray}
  &&\delta_4 \psi_1 = \, 2 g \, \eta_4 \, [\Phi^1,\Phi^2] \,,
\qquad (\delta_4)^2 \psi_1 = 0 \nonumber  \, , \\
  &&\delta_4 \Phi^3 = \eta_4 \, \psi_1 \,, \qquad (\delta_4)^2 \Phi^3 = \,
2 g \, \eta_4^2 \, [\Phi^1,\Phi^2] \label{27} \, , \\
  && \delta_4 \Phi^1 = \delta_4 \Phi^2 = 0 \,, \nonumber
\end{eqnarray}
we can immediately obtain eqs. (\ref{susyBF}).

\subsubsection{The operator $\cK_{\mathbf{6}}$}\label{subL6}

Finally, here are the relevant expressions for the BMN operator $\cK_{\mathbf{6}}$. The primary state of the multiplet is in the \textbf{6} of $SU(4)$ with $\cN=1$ (i.e., $SU(3)\times U(1)$) projection  $\cK^I_{\mathbf{6}} \, = \, \T(\Phi^I \Phi^J \bar\Phi_J) + \T(\Phi^I \bar \Phi_J \Phi^J)$.
It has a descendant $\cK_{\mathbf{45}}$ in the \textbf{45} (DL $[210]$) whose highest weight all-chiral projection is a mixture of the operators\footnote{We use the same notation as in the Konishi case, e.g., $B$, $F$, etc., although the meaning is now different.}
\begin{equation}
B \, = \, \T(\Phi^1 \Phi^1 [\Phi^2, \Phi^3]) \, , \qquad  F \, = \,
\T(\Phi^1 \, W^\alpha \, W_\alpha) \, .
\end{equation}
The right mixture can be determined by orthogonalization since, in complete analogy with the
Konishi case, only the $B$ term follows from the classical field equations whereas
the $F$ part is a generalized ``anomaly". Notice the important difference with the Konishi case: Even in the free theory, where $\cK_{\mathbf{6}}$ satisfies a semishortness condition of the type (\ref{BMNsemi}), it does not contain any conserved current.

Performing two supersymmetry transformations with parameter $\eta_4$ (recall (\ref{27})) we find
\begin{equation}
(D^4)^2 \, B \, = \, g \, Y \, , \qquad
(D^4)^2 \, F \, = \, 2 g^2 \, Y \label{susyBF'}
\end{equation}
with the bare descendant
\begin{equation} \label{y300}
Y \, = \, - 2 \, \T(\Phi^1 \Phi^1 [\Phi^1, \Phi^2] \Phi^2) \, ,
\end{equation}
which is the highest weight of a \textbf{300} (DL $[212]$) \cite{Ryzhov:2001bp}.

The same transformation applied to the mixture
\begin{equation}
\cO_{\mathbf{45}} \, = \, F \, - \, 2 g \, B
\end{equation}
gives zero. This combination can be identified as a descendant of the protected half-BPS multiplet $\cO_{\mathbf{50}}$ (DL $[030]$).

\section{Calculation of $\gamma_3$ for the Konishi multiplet}\label{s3}

\subsection{The setup}\label{s31}

In subsection \ref{ss1} we showed that the two ratios of two-point functions (\ref{ratio1}) and (\ref{ratio2}) contain the same information about the anomalous dimension of the Konishi operator. The presence of an overall factor $g^2$ in both sides of these equations means that in order to determine, e.g., $\gamma_3$, we need to compute one of these ratios at level $g^4$ (``two loops"). We also explained why we choose to evaluate the second ratio,
\begin{equation}\label{R}
R \, = \, \frac{\langle \, \cK_{\mathbf{84}} \, \cK_{\mathbf{84}}^\dagger \,
\rangle}{\langle \, \cK_{\mathbf{10}} \, \cK_{\mathbf{10}}^\dagger \, \rangle}\ .
\end{equation}
Once the ratio (\ref{R}) is expanded in powers of $g^2$, prior to the
evaluation of any Feynman diagram, we will see that similar linear combinations
of integrals occur in both two-point functions which leads to substantial
cancellations. The remaining set of integrals can be drastically reduced by
comparison with the two-point function of a certain half-BPS operator. The
latter is known to be non-renormalized, so the corresponding set of Feynman
graphs must sum into contact terms.

We begin by recalling eq. (\ref{13'}) which states that the renormalized operator $\cK_{\mathbf{10}}$ is a mixture of the bare operators $B$ and $F$ with some singular renormalization factors. In the supersymmetric dimensional reduction scheme these factors have the following form \cite{us}
\begin{eqnarray}
Z_B & = & 1 \, + \, g^2 \, \frac{b_{11}}{\epsilon} \, + \, g^4 \,
\left( \frac{b_{22}}{\epsilon^2} \, + \, \frac{b_{21}}{\epsilon}
\right) \, + \, O(g^6) \, , \\ Z_F & = & g \, f_{10} \, + \, g^3 \,
\left( \frac{f_{21}}{\epsilon} \, + \, f_{20} \right) \, + \,
O(g^5) \, .
\end{eqnarray}
An overall finite rescaling of the operator $\cK_{\mathbf{10}}$ carries over to its
descendant  $\cK_{\mathbf{84}}$ and drops in the ratio of two-point
functions (\ref{R}). In the above we used this freedom to remove the finite part of $Z_B$. However, $Z_F$ must contain a finite part, the terms $f_{10}$ and $f_{20}$ ($f_{10}$ is the ``one-loop" Konishi anomaly). In \cite{us} it is shown that $f_{21}$ as well as all the $b_{mn}$ are determined by the anomalous dimension of the Konishi multiplet. In our calculations here, as shown below, the coefficients $b_{22}$, $b_{21}$ and $f_{20}$ drop out, so we only need $b_{11}$ and $f_{21}$, which will be determined directly.

Let us combine (\ref{susyBF}) and (\ref{susyBF'}) into one equation suitable for treating both cases,
\begin{equation}
D^4 \cdot D^4 \, B \, = \, g \, Y \qquad D^4 \cdot D^4 \, F
\, = \, - a g^2 \, Y \label{delQ}\,,
\end{equation}
where $a=-4$ in the case of $\cK_{\mathbf{1}}$ and $a=-2$ for
$\cK_{\mathbf{6}}$. (By definition $a$ is the coefficient in the protected
linear combination $F + a g \, B$.)

As discussed earlier, we assume that the quantum operator $\cK_{\mathbf{84}}$ is obtained from $\cK_{\mathbf{10}}$ by naive application of the field equations:
\begin{equation}
g\, \cK_{\mathbf{84}} \, = \, D^4 \cdot D^4 \, \cK_{\mathbf{10}} \, = \,
g \, \left( Z_B \, - \, a g \, Z_F \right) \, Y \, = \, g \, Z_{\mathbf{84}} \,
Y
\end{equation}
Explicitly,
\begin{equation}
Z_{\mathbf{84}} \, = \, 1 \, + \, g^2 \, \left( \frac{b_{11}}{\epsilon} \,
- \, a \, f_{10} \right) \, + \, g^4 \, \left(
\frac{b_{22}}{\epsilon^2} \, + \, \frac{b_{21} - a f_{21}
}{\epsilon} \, - \, a \, f_{20} \right) \, + \, O(g^6) \,.
\end{equation}
The detailed analysis of the two-point functions below confirms that this
definition of $Z_{\mathbf{84}}$ renders the operator $\cK_{\mathbf{84}}$
finite. \newline

Before going on, in order to check our normalizations, let us show how the
well-known value of $\gamma_1$ is obtained from the ratio (\ref{R}). At the
lowest order we have the equation
\begin{equation}\label{28}
   \frac{g^2\, \langle \, \cK_{\mathbf{84}} \, \cK_{\mathbf{84}}^\dagger \,
\rangle_{g^0}}{\langle \, \cK_{\mathbf{10}} \, \cK_{\mathbf{10}}^\dagger \, \rangle_{g^0}} = \frac{g^2 \, YY_0}{BB_0} = - \frac{2 g^2 \gamma_1}{x^2_{12}}\,.
\end{equation}
Here
\begin{eqnarray}
  BB_0 & = & \langle \, B \, B^\dagger \, \rangle_{g^0} =
- \frac{2N(N^2-1)}{(4 \pi^2 x_{12}^2)^3} \nonumber \, , \\
  YY_0 & = & \langle \, Y \, Y^\dagger \, \rangle_{g^0}  =  \frac{12N(N^2-1)}
{(4 \pi^2 x_{12}^2)^4} \,   \label{29}
\end{eqnarray}
are the tree-level two-point functions of the scalar operators $B$ and $Y$ (see Section \ref{s331}). It follows
\begin{equation}\label{30}
  \gamma_1 = \frac{3N}{4\pi^2}\, ,
\end{equation}
as first computed in \cite{Anselmi:1996mq}.

\subsection{Evaluating the two-point functions}\label{s32}

Now we need to evaluate the two-point functions of the renormalized operators $\cK_{\mathbf{10}}$ and $\cK_{\mathbf{84}}$ up to order $g^4$. In the process we encounter the various two-point functions of the bare operators $B$ and $F$. We introduce the notation
\begin{eqnarray}
BB_2
& = & \frac{\langle \, B \, B^\dagger \, \rangle_{g^2}}{g^2 \,
BB_0} \, , \nonumber \\ BB_4 & = &  \frac{\langle \, B \,
B^\dagger \, \rangle_{g^4}}{g^4 \, BB_0} \, , \nonumber \\ BF_3 &
= & \frac{\langle \, B \, F^\dagger \, \rangle_{g^3}}{g^3 \, BB_0}
\, , \nonumber \\ FF_0 & = & \frac{\langle \, F \, F^\dagger \,
\rangle_{g^0}}{BB_0} \, , \nonumber \\ FF_2 & = & \frac{\langle \,
F \, F^\dagger \, \rangle_{g^2}}{g^2 \, BB_0} \, , \nonumber \\
 YY_2
& = & \frac{\langle \, Y \, Y^\dagger \, \rangle_{g^2}}{g^2 \,
YY_0} \, , \nonumber \\ YY_4 & = & \frac{\langle \, Y \, Y^\dagger
\, \rangle_{g^4}}{g^4 \, YY_0} \, . \nonumber
\end{eqnarray}
We remark that $\langle \, B \, F^\dagger \, \rangle_{g^1} \, = \,
0$ simply because no such graph exists. For the two-point functions of the
renormalized operators we find
\begin{eqnarray}
\frac{\langle \, \cK_{\mathbf{10}} \, \cK_{\mathbf{10}}^\dagger \rangle}{BB_0} & = &
Z_B^2 \, + \, BB_2 \, \left(g^2 + g^4 \frac{2 b_{11}}{\epsilon}
\right) \, + \, BB_4 \, g^4 \, + \, BF_3 \, g^4 \, 2 f_{10}
\label{exp10} \\ && + \, FF_0 \, \left(g^2 \, f_{10}^2 + g^4(
\frac{2 f_{10} f_{21}}{\epsilon} + 2 f_{10} f_{20} ) \right) \, +
\, FF_2 \, g^4 \, f_{10}^2 \, + \, O(g^6) \nonumber
\end{eqnarray}
and
\begin{eqnarray}
\frac{\langle \, \cK_{\mathbf{84}} \, \cK_{\mathbf{84}}^\dagger
\rangle}{YY_0} & = & Z_B^2 - g^2 \, 2 a f_{10} - g^4 \, 2 a \left(
\frac{b_{11} f_{10} + f_{21}}{\epsilon} + f_{20} - \frac{a}{2}
f_{10}^2 \right) \label{exp84} \\ && + \, YY_2 \, \left( g^2 \, +
g^4 \, (\frac{2 b_{11}}{\epsilon} - 2 a f_{10}) \right) \, + \,
YY_4 \, g^4 \, + \, O(g^6) \nonumber \, .
\end{eqnarray}
We have normalized such that in both equations the r.h.s. goes
like $1 + O(g^2)$. The ratio of (\ref{exp84}) and (\ref{exp10})
can therefore straightforwardly be expanded in the coupling
constant:
\begin{eqnarray}
\frac{BB_0}{YY_0} \; R & = & 1 \, + \, g^2
\left( (YY_2 - BB_2) - f_{10} (a + (a + f_{10} \, FF_0)) \right)
\label{master1} \\ && + \, g^4 \, \left( YY_4 - BB_4 - BB_2 \,(
YY_2 - BB_2 ) \right) \nonumber \\ && - \, g^4 \, f_{10} \left( 2
\, BF_3 + a \, YY_2 \right) - g^4 \, \frac{2 f_{21}}{\epsilon} (a +
f_{10} \, FF_0) \nonumber
\\ && + \, g^4 \, \left( 2(BB_2 + \frac{2 b_{11}}{\epsilon}) -
(YY_2 + \frac{2 b_{11}}{\epsilon}) \right) \, f_{10} \, (a + f_{10}
\, FF_0) \nonumber
\\ && - \, g^4 \, \left(2 f_{20} - f_{10}^2 (a
+ f_{10} \, FF_0) \right)(a + f_{10} \, FF_0) \, - \, g^4 \,
f_{10}^2 \, FF_2 \nonumber
\end{eqnarray}
Note that the $g^4$ part of $Z_B$ has dropped out.

Let us try to
simplify further. In Section \ref{s23} we explained that the combination
\begin{equation}
\cO_{\mathbf{10}} \, = \, F \, + \, a g \, B
\end{equation}
(likewise $\cO_{\mathbf{45}}$ for the $J=1$ case) is short and thus
protected. In particular (see (\ref{21})),
\begin{equation}
\langle \, \cO_{\mathbf{10}} \, \cO_{\mathbf{10}}^\dagger \, \rangle_{g^2} \, = \, 0
\end{equation}
up to contact terms. Since $\langle B F^\dagger \rangle_{g^1} = 0$ we can
deduce
\begin{equation}
FF_2 = - a^2 \label{elimFF2}
\end{equation}
up to a contact term. In dimensional regularization, two-point
contact terms are $O(\epsilon)$ as long as the points are kept
apart \cite{Penati:2000zv}. For our purposes we can put them to zero provided that they
are not multiplied by a singular term coming, e.g., from the $Z$ factors.
In (\ref{master1}) $FF_2$ appears with a finite factor and thus it
can be replaced by $-a^2$.

Next, we recall that the operators  $\cO_{\mathbf{10}}$ and $\cK_{\mathbf{10}}$ must be orthogonal, see (\ref{22}). To first order in $g$ this means
\begin{equation}\label{30'}
\langle (F + a g \, B) (B^\dagger + f_{{10}} g \, F^\dagger) \rangle_{g^1} \, = \, g \, BB_0 ( a \, + \, f_{10} \, FF_0 ) \, = \, O(\epsilon)\,,
\end{equation}
where we have again used $\langle B F^\dagger \rangle_{g^1} = 0$. Eq. (\ref{30'}) implies
 \begin{equation}
a + f_{10} \, FF_0 \, = \, O(\epsilon) \, .
\label{elimFF0}
\end{equation}
Notice that in dimensional regularization the right-hand side of eq. (\ref{30'}) is not identically zero. The orthogonality condition becomes exact only in the limit $\ep\to 0$. While we
try to drop the expression $a + f_{10}\, FF_0$ wherever possible,
we cannot do so when it occurs with a singular factor. For instance, the
term $\frac{f_{21}}{\epsilon} \, (a + f_{10} \, FF_0)$ must be kept.

The terms in the fourth line of eq. (\ref{master1}) can be dropped since the combinations
\begin{equation}
BB_2 + \frac{2 b_{11}}{\epsilon} \, = \, O(1) \, = \, YY_2 +
\frac{2 b_{11}}{\epsilon} \label{renBBYY}
\end{equation}
are finite. Indeed, the poles $\ep^{-1}$ in both $BB_2$ and $YY_2$ have residues equal to their common anomalous dimension $\gamma_1$. So, these poles are removed by the same one-loop renormalization factor with
$b_{11} \, = \, \gamma_1 / 2$, as shown in (\ref{renBBYY}) (see also (\ref{BBYYAC}) below).

Further,  the difference $YY_2 - BB_2$ in the first line of eq. (\ref{master1}) can be dropped as well. Indeed, both $BB_2$ and $YY_2$ contain only two graphs which are denoted by A and C in Figure 1. In Appendix \ref{3.1} it is shown
that Graph C is a contact term. The operators $B$ and $Y$ have the same poles
$\ep^{-1}$ (and hence the same anomalous dimension $\gamma_1$) exactly
because diagram A occurs in $YY_2$ and $BB_2$ with the same coefficient,
see (\ref{BBYYAC}) below. Thus,
\begin{equation}
YY_2 - BB_2 \, = \, O(\epsilon) \, . \label{elimBBYY}
\end{equation}
Notice, however, that the same difference cannot be neglected in the second line of eq. (\ref{master1}) because it is multiplied by the divergent two-point function $BB_2$.

Putting together eqs. (\ref{elimFF2}) - (\ref{elimBBYY}), we can considerably
simplify eq. (\ref{master1}):
\begin{eqnarray}
\frac{BB_0}{YY_0} \; R & = & 1 \, - \, a \,
f_{10} \, g^2 \, + \, (a \, f_{10} \, g^2)^2 \label{master2} \\ &&
+ \, g^4 \, \left( YY_4 - BB_4 - BB_2 \,( YY_2 - BB_2 ) \right)
\nonumber \\ && - \, g^4 \, f_{10} \left( 2 \, BF_3 + a \, YY_2 \right)
- g^4 \, \frac{2 a f_{21}}{\epsilon} (1 + \frac{f_{10}}{a} \,
FF_0) \nonumber
\end{eqnarray}

We can still simplify the last term in (\ref{master2}) in a fashion that is universal to the $J=0,1$ cases:
\begin{eqnarray}\label{33}
BB_0 & \propto & \Pi_{12}^{(J+3)} \, , \\ \langle \, F \, F^\dagger
\, \rangle_{g^0} & \propto & \Pi_{12}^J (\partial \Pi_{12})^2
\nonumber \, .
\end{eqnarray}
(For the exact expressions see the next subsection.)
The configuration space propagator in dimensional regularization is
\begin{equation}\label{propdimreg}
\Pi_{12} \, = \, - \frac{\Gamma(1-\epsilon) \pi^\epsilon}{4 \pi^2
(x_{12}^2)^{(1-\epsilon)}} \,
\end{equation}
Now, the requirement of orthogonality (\ref{elimFF0}) implies
\begin{equation}
\frac{f_{10}}{a} \, \frac{\langle \, F \, F^\dagger \,
\rangle_{g^0}}{BB_0} \, = \, \frac{f_{10}}{a} \, FF_0 \, = \, - 1
+ O(\epsilon) \, .
\end{equation}
The $\ep$ dependence in this expression comes from the expansion of the
propagators (\ref{propdimreg}).  In the $\overline{MS}$ scheme the
fractional powers of $\pi$ as well as the Euler-Mascheroni constant from
the expansion of $\Gamma$-functions are absorbed into the mass
scale, see Appendix \ref{AppC}. In this scheme
\begin{equation}\label{sudden}
\frac{f_{10}}{a} \, FF_0 \, = \, - (1-\epsilon)^2 (x^2_{12})^{(-\epsilon)} \,
(1 - \epsilon^2 \frac{\zeta(2)}{2} + \ldots \, )
\end{equation}
and finally
\begin{equation}
\frac{1}{\epsilon} \, \left(1 \, + \, \frac{f_{10}}{a} \, FF_0
\right) \, = \, \ln( x_{12}^2 ) \, + \, 2 \, + \, O(\epsilon)
\, .
\end{equation}

\subsection{Evaluation of the graphs}\label{s33}

In this section we elaborate the contributions $BB_2$, $YY_2$, $BF_3$
and $YY_4-BB_4$ which occur in (\ref{master2}).

\subsubsection{Level $g^0$}\label{s331}

At tree level we have
\begin{equation}
BB_0 \, = \langle \, B \, B^\dagger \, \rangle_{g^0} \, = \,
\langle \, (\Phi^1 [\Phi^2, \Phi^3]) \, ([\bar \Phi_3, \bar
\Phi_2] \bar \Phi_1) \, \rangle \, = \, 2 N (N^2-1) \, \Pi_{12}^3
\, .
\end{equation}
One comment on the convention: the linear part of the $W^\alpha$
field is
\begin{equation}
W^\alpha \, = \, - \frac{1}{4} \bar D^2 D^\alpha V \, .
\end{equation}
Its lowest component is the physical fermion $\lambda^\alpha$ of
the $\cN=1$ SYM multiplet. One may verify from the definitions in Appendix
\ref{3.1} that indeed
\begin{equation}
\langle \, W_\alpha(1) \, \bar W_{\dot \alpha}(2) \, \rangle_{g^0,
\,  \theta, \bar\theta = 0} \, = \, - 2 i \, \partial_{\alpha \dot
\alpha}|_1 \Pi_{12}|_{\q, \bq \, = \, 0}
\end{equation}
in agreement with our convention for the matter fermion
propagator. Hence we find
\begin{equation}
\langle \, F \, F^\dagger \, \rangle_{g^0} \, = \, - 16 (N^2-1) \,
(\partial_\mu \Pi_{12})^2 \, .
\end{equation}
We can now use the orthogonalization condition (\ref{elimFF0}) to
fix the one-loop anomaly coefficient $f_{10}$. In $\cO_{\mathbf{10}}$ we have
$a=-4$ so that
\begin{equation}
f_{10} \, = \, \frac{N}{32 \pi^2} \, .
\end{equation}
The remaining tree-level correlator is
\begin{equation}
YY_0 \, = \, \langle \, Y \, Y^\dagger \, \rangle_{g^0} \, = \, 12
N^2 (N^2-1) \, \Pi_{12}^4 \, .
\end{equation}

\subsubsection{Level $g^2$}

We now turn our attention to the $O(g^2)$ contributions $BB_2$ and $YY_2$.
Here and in the sequel we ignore the one-loop self-energy diagrams since
they cancel in the $\cN=1$ formulation of ${\cal N}=4$ SYM. Further, we do not
display free matter lines.

\vskip 25pt
\noindent
\begin{minipage}{\textwidth}
\begin{center}
\includegraphics[width=0.70\textwidth]{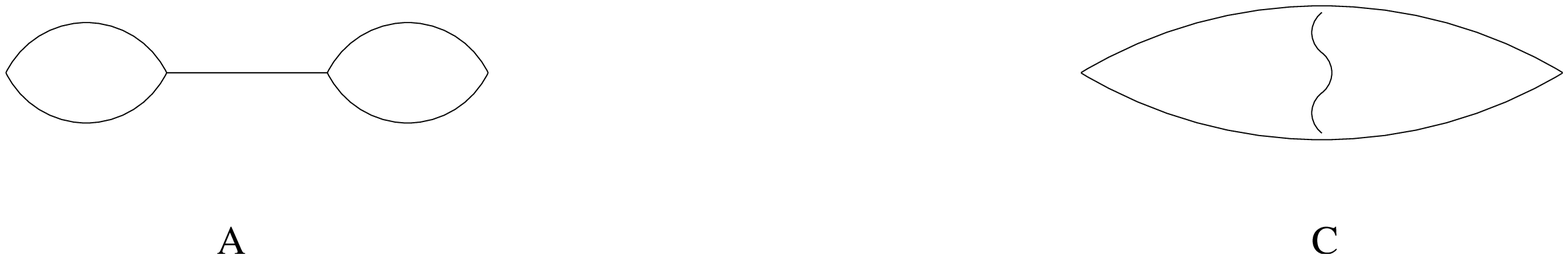}
\end{center}
\end{minipage}
\begin{center}
Figure 1. Graphs contributing to $BB_2$ and $YY_2$; the free matter lines are not shown.
\end{center}
\vskip 15pt

We find
\begin{eqnarray}
\langle \, B \, B^\dagger \, \rangle_{g^2} & = & 6 g^2 N^2 (N^2-1)
\, (2 \, \Pi_{12} \, A \, + \Pi_{12} \, C) \, , \label{BB2YY2} \\
\langle \, Y \, Y^\dagger \, \rangle_{g^2} & = & 36 g^2 N^3 (N^2-1)
\, (2 \, \Pi_{12}^2 \, A \, + \, \frac{4}{3} \, \Pi_{12}^2 \, C)
\, , \nonumber
\end{eqnarray}
with the graphs from Figure 1, which are calculated in Appendix \ref{3.1}.
Note that in our conventions the YM propagator has a minus sign
relative to the matter propagator $\Pi_{12}$. This sign has
already been taken into account in the combinatorial factors in
(\ref{BB2YY2}). On scaling down by the respective tree-levels, we obtain
\begin{eqnarray}
BB_2 & = & 3 N \, (2 \, \frac{A}{\Pi_{12}^2} \, + \,
\frac{C}{\Pi_{12}^2}) \, , \label{BBYYAC} \\ YY_2 & = & 3 N \, (2
\, \frac{A}{\Pi_{12}^2} \, + \, \frac{4}{3} \,
\frac{C}{\Pi_{12}^2}) \, . \nonumber
\end{eqnarray}
This confirms the claim made in (\ref{elimBBYY}).

\subsubsection{Level $g^3$ \label{g3s}}

Next, we turn our attention to $\langle \, B \, F^\dagger \, \rangle_{g^3}$. We recall that the operators $B$ and $F$ are highest weight projections of the \textbf{10} of $SU(4)$. It proves more convenient to calculate the two-point function of a different projection of the
\textbf{10}, namely $B' = (\Phi^1 [\bar \Phi_2, \bar \Phi_3])$ and
$F' = ((\nabla^\alpha \Phi^1)(\nabla_\alpha \Phi^1))$. One sees
immediately that $\langle \, B \, B^\dagger \, \rangle_{g^0} \, =
\, \langle \, B' \, {B'}^\dagger \, \rangle_{g^0}$, i.e., the
normalization of the two components is the same. For $F$ and $F'$
we have checked this in the beginning of the section. Thus we can
safely appeal to $SU(4)$ and generalize results found for this
projection of the \textbf{10} to the entire $SU(4)$ multiplet.\footnote{A word of caution is due here. Part of the $SU(4)$ R symmetry may turn out anomalous in the quantum theory. However, the transformation leading from $B$ to $B'$ and from $F$ to $F'$ actually uses the $SU(2) \subset SU(4)$ which can be maintained manifest in the off-shell $\cN=2$ harmonic superspace \cite{Galperin:1984av} formulation, see \cite{Arutyunov:2002jg}.}

The operator $F'$ contains the YM covariant derivative
\begin{equation}
\nabla^\alpha \, \Phi^1 \, = \, D^\alpha \, \Phi^1 \, - \, g \, [ D^\alpha V,
\Phi^1 ] \, + \, O(g^2)
\end{equation}
and $B'$ is made gauge invariant by insertion of the gauge bridge $e^{gV}$
between superfields of different chirality.

To order $g^3$ there are three types of diagrams: First, those
that do not involve the gauge connection at the $F'$ point. These are
the topologies $G_{I}$
through $G_{III}$ in Figure 2 and a few other graphs that vanish when
$\theta, \bar\theta \, = \, 0$ at the outer points. Second, we may
have one connection line coming out of $F'$. There are six
related graphs, of which the only non-zero one is displayed in
Figure 2 as graph $G_{IV}$. Third, there are also graphs with a single
YM line emanating from the gauge bridge in $B'$. However, they do
not contribute.

\vskip 20pt
\noindent
\begin{minipage}{\textwidth}
\begin{center}
\includegraphics[width=0.70\textwidth]{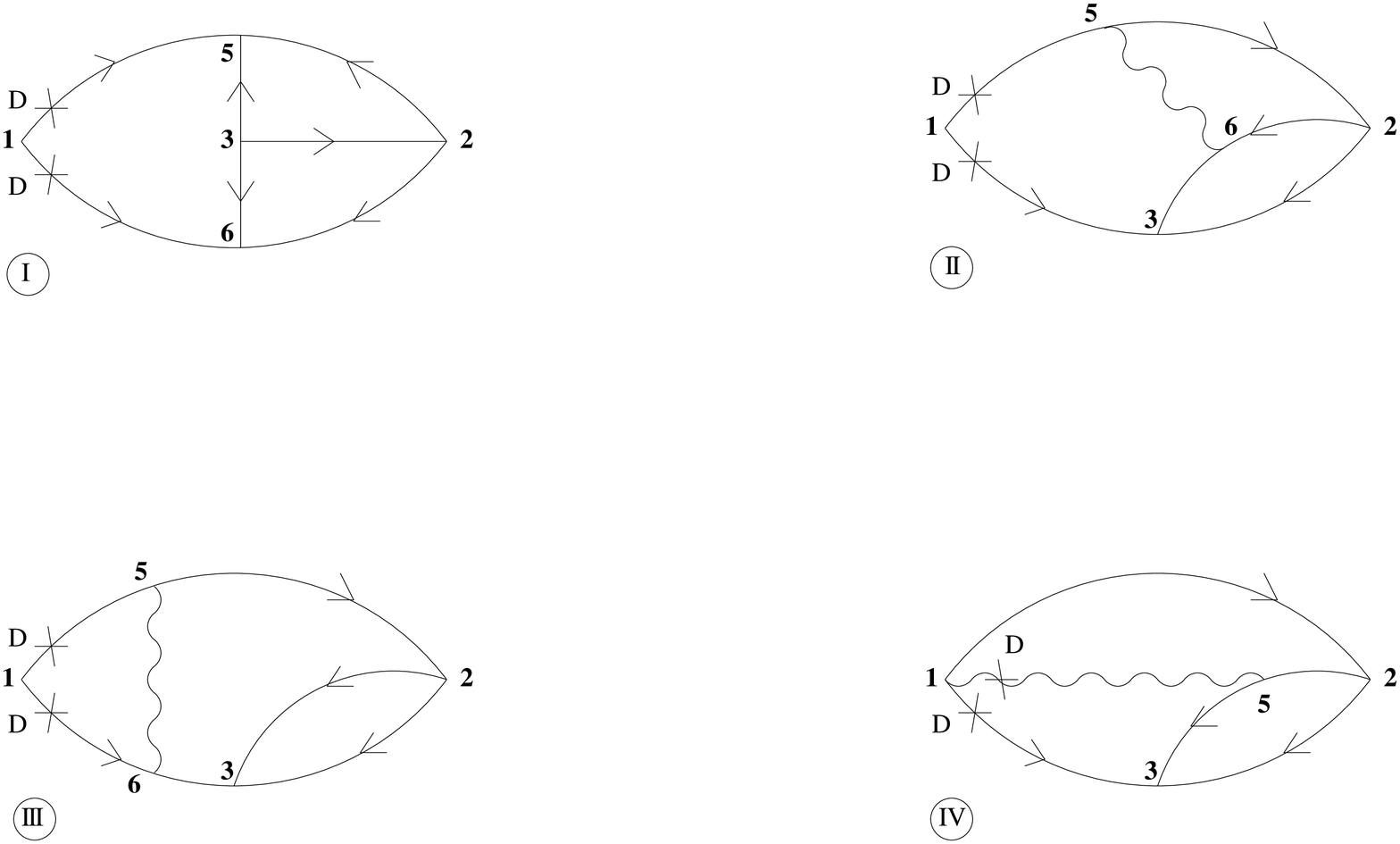}
\end{center}
\end{minipage}
\begin{center}
Figure 2. Graphs $G_{I\cdots IV}$; the crosses denote spinor derivatives.
\end{center}
\vskip 15pt

The combinatorics for the non-vanishing graphs yields the sum
\begin{equation}
\langle \, F' \, {B'}^\dagger \, \rangle_{g^3} \, = \, - 8 g^3 N^2
(N^2-1) \, \left( \frac{1}{2} \; G_{I} \, - \, G_{II} \, + \, G_{III} \, -
G_{IV} \right) \, ,
\end{equation}
where the minus sign from the YM propagator is included. The signs
in the latter formula are valid when the propagator with
$D^\alpha$ is written to the left of the one with $D_\alpha$ and when
in $G_{IV}$ the $D^\alpha$ is on the YM line. The Grassmann integrations
in all four supergraphs are straightforward; in the YM sector we additionally
assemble double derivatives into box operators to find
\begin{equation}
G_{I} \, = \, - G_{II} \, + \, G_{III} \, - \, G_{IV} \, = \, 2 \, J \, ,
\end{equation}
and hence
\begin{equation}
\langle F \, B^\dagger \rangle_{g^3} \, = \, - 24 g^3 N^2 (N^2-1) \;
J \, .
\end{equation}
The letter $J$ stands for the Feynman integral (the notation is explained
in the Appendix)
\begin{equation}
J \, = \, c_0^7 \, \int \frac{d^{4-2\ep}x_{3,5,6} \, (\partial_{15} \,
\partial_{53} \, \partial_{63} \, \partial_{16})}{(x_{15}^2 \,
x_{16}^2 \, x_{35}^2 \, x_{36}^2 \, x_{25}^2 \, x_{23}^2 \,
x_{26}^2)^{(1-\epsilon)}} \, .
\end{equation}
(The factor $c_0 \, = \, - \Gamma(1-\epsilon) \pi^\epsilon / (4
\pi^2)$ comes from the normalization of the configuration space
propagator.)

The evaluation of $J$ can be done by switching over to momentum space and using
the powerful {\it Mincer} computer algorithm \cite{Mincer}. $J$ is a three-loop
integral of topology \textbf{BU} in the {\it Mincer} classification. The
program computes
\begin{equation}
J \, = \, \frac{1}{(4 \pi)^6} \, \frac{1}{\epsilon^2} \, \left(
\frac{1}{2} \, + \, \frac{14}{3} \, \epsilon \, + \, O(\epsilon^2)
\right) \, (q^2)^{(1 - 3 \epsilon)} \,
\end{equation}
in the $\overline{MS}$ scheme. We transform back to configuration
space:
\begin{equation}
J \, = \, - \frac{1}{16 \pi^2} \, \left(\frac{3}{\epsilon} \, + \, 1
\, + \, O(\epsilon) \right) \, \frac{1}{(4 \pi^2 x_{12}^2)^3} \,
(x_{12}^2)^{(4 \epsilon)}
\end{equation}
For the definitions of the Fourier transform and the mass scales
in $p$- and $x$-space see Appendix \ref{AppC}. Finally, since $J$ is real:
\begin{equation}
BF_3 \, = \, - \frac{3 N}{4 \pi^2} \, \left( \frac{3}{\epsilon} \, +
\, 1 \, + \, O(\epsilon) \right) \, (x_{12}^2)^\epsilon
\end{equation}

Next, we turn to $YY_2$, which we need up to $O(\epsilon)$.
Graph C is of no interest in this context as it is a contact term. Graph A
is computed in detail in Appendix \ref{3.1}, equation (\ref{evA}). On
dividing by the tree-level $BB_0$ we obtain
\begin{equation}
YY_2 \, = \, - \frac{3 N}{4 \pi^2} \, \left( \frac{1}{\epsilon} \,
+ \, 1 \, + \, O(\epsilon) \right) \, (x_{12}^2)^\epsilon \, .
\end{equation}
The one-loop anomalous dimension of the operator $Y$ (and in the same
way of $B$) is therefore $\gamma_1 = 3 N / (4 \pi^2)$, as stated earlier (see (\ref{renBBYY})). \newline

Collecting terms:
\begin{equation}\label{oncol}
- \, f_{10} \, ( 2 \, BF_3 \, - \, 4 \, YY_2 ) \, = \, \frac{3}{4} \, \frac{
N^2}{(4 \pi^2)^2} \,  \left( \frac{1}{\epsilon} \, - \, 1 \, + \, O(\epsilon)
\right) \, (x_{12}^2)^\epsilon
\end{equation}

\subsubsection{Level $g^4$}\label{334}

It might seem that the evaluation of the genuine four-loop
integrals in $YY_4$ and $BB_4$ is an insurmountable obstacle - we
are now going to show how to get away with calculating only one
difficult integral.

The $O(g^4)$ diagrams that occur in $YY_4$ and $BB_4$ are listed in
Figure 3. One of the advantages of the all-chiral choice for $Y$
and $B$ is that the gauge bridge $e^{g V}$ does not occur.
Therefore all diagrams have the same number of outer lines at
both ends, namely four lines in $YY_4$ and three in $BB_4$. Also,
all the outer lines connect to matter fields. Now, any graph has one or two connected
pieces and a number of free lines. In Figure 3 we do not show such
``spectator propagators". Also, we do not display the ``mirror" graphs obtained by permuting points 1 and 2; if needed, they will be denoted by, e.g., $G_{17b}$ (see (\ref{tidy})).

Above we have first computed $BF_3$ and $YY_2$ and then divided by
the respective tree-levels. Here we divide out the tree-level
before calculation - in practical terms this means first of all to
divide the combinatorial coefficient of any graph in $YY_4$ by $12
N^2 (N^2-1)$ and that of any graph in $BB_4$ by $2 N (N^2-1)$.
Further, we ``normalize" the integrals by $\Pi_{12}^4$ and
$\Pi_{12}^3$, respectively. In doing so we may cancel spectator
lines, because the propagator is non-singular. As a result, an
integral with $n$ outer legs at each end will occur with $n$
inverse powers of $\Pi_{12}$.

\vskip 20pt
\noindent
\begin{minipage}{\textwidth}
\begin{center}
\includegraphics[width=0.70\textwidth]{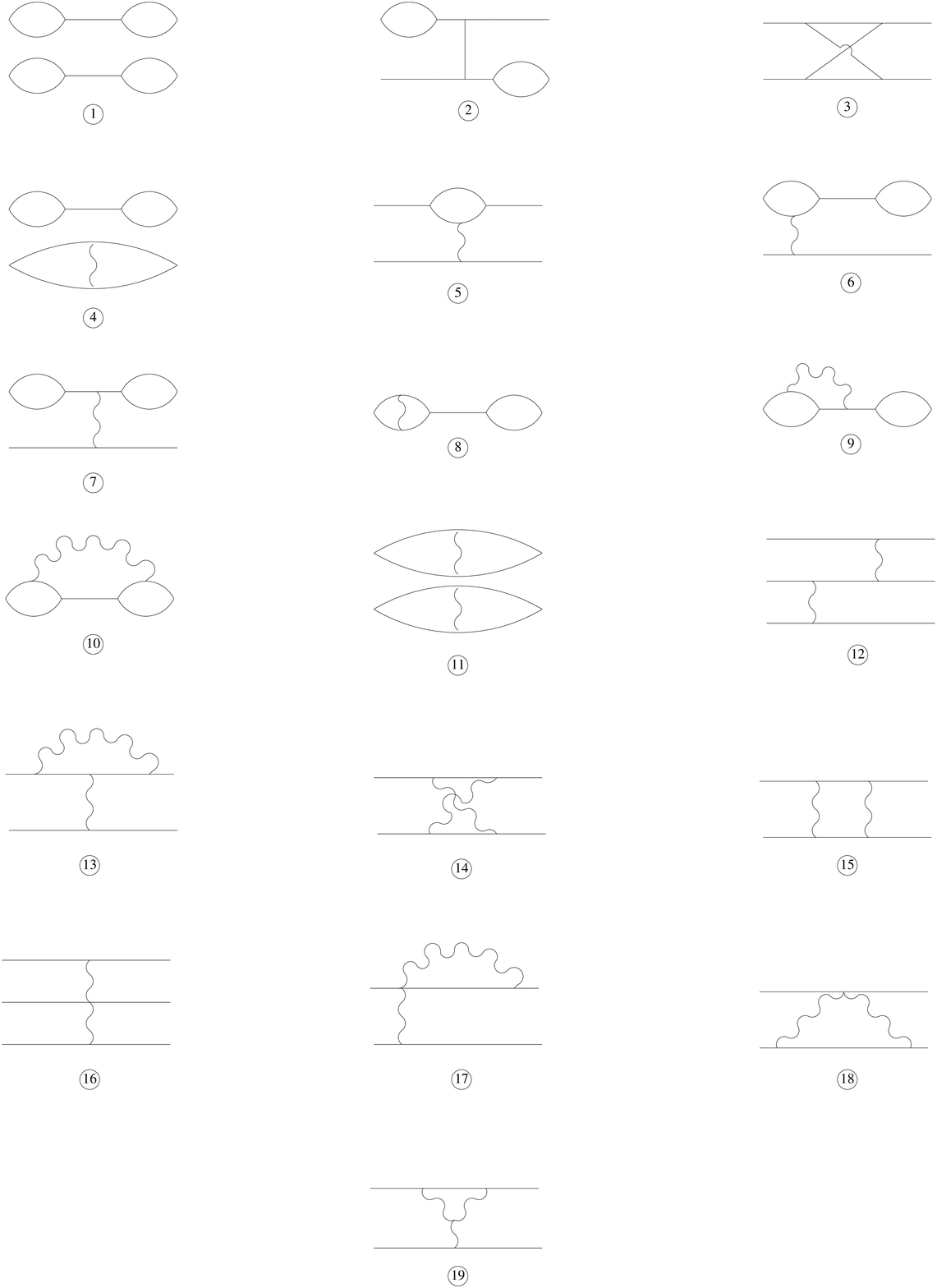}
\end{center}
\end{minipage}
\begin{center}
Figure 3. Graphs $G_{1\cdots 19}$; the mirror graphs are not shown; the outer points are split for clarity.
\end{center}
\vskip 15pt

Apart from the 19 graphs in the list the correlators also contain
six further diagrams that together constitute the two-loop
propagator correction. The propagator blob occurs once per matter
line, so once more in $YY_4$ than in $BB_4$.\footnote{We do not
argue in this paper whether the two-loop blob is just finite or
really vanishing. This is presumably a scheme dependent issue;
here we do not need the explicit result.}

\vskip 20pt
\noindent
\begin{minipage}{\textwidth}
\begin{center}
\includegraphics[width=0.70\textwidth]{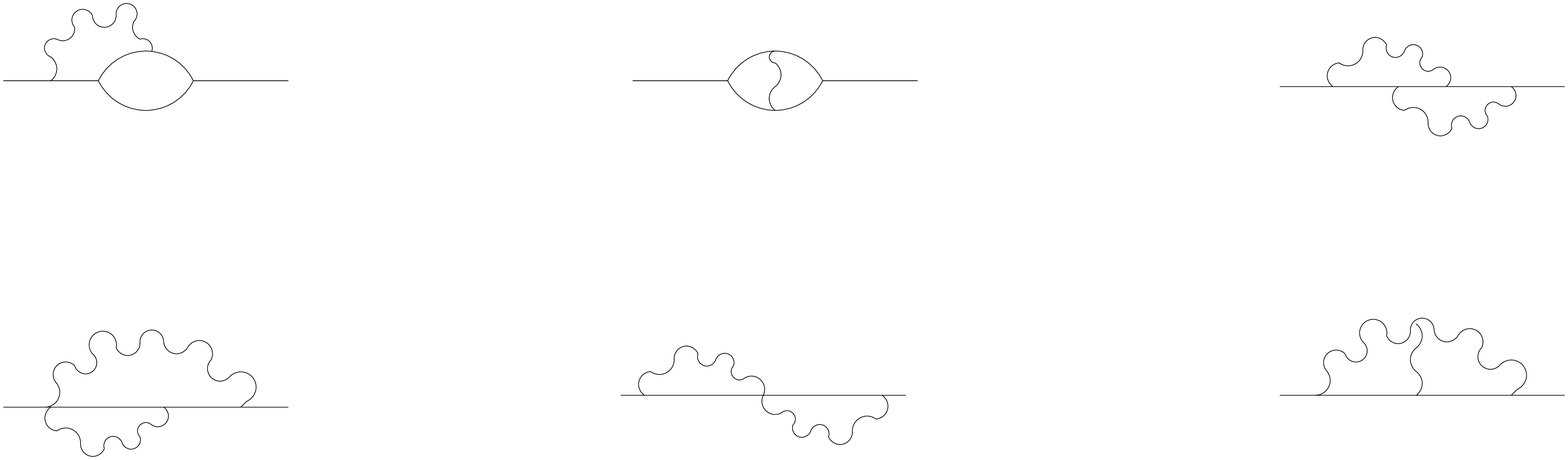}
\end{center}
\end{minipage}
\begin{center}
Figure 4. Propagator correction graphs denoted by $ -\bigcirc-$ in the text.
\end{center}
\vskip 15pt

To be more tidy, we directly give the difference $YY_4 \, - \,
BB_4$ in which graphs $G_{6}$ through $G_{10}$ drop:
\begin{equation}
YY_4 \, - \, BB_4 \, = \, 12 N^2 \, \left( \frac{1}{2} \, G_{1} \, +
\, \frac{1}{2} \, G_{4} \, + \, \frac{1}{8} \, G_{11} \, - \, G_{2} \right)
\, + \, 2 N^2 \, \Sigma_1 \, + \, 24 \, \Sigma_2
\end{equation}
The two sums of diagrams are
\begin{eqnarray}
\Sigma_1 & = & - \, 2 \; \, G_{5} \, + \, \frac{1}{4} \, G_{11} \, + \,
G_{12} \, - \, G_{13} \, + \, \frac{1}{2} \, G_{15} \, + \,
G_{16} \label{tidy}\\
&& - \, 3 \, (G_{17} \, + \, G_{17b}) \, + \, G_{18} \, + \, 2
\; \, G_{19} \, + \, \frac{1}{2 N^2 g^4} \, \left[ -\bigcirc-\right] \nonumber \, ,
\end{eqnarray}
where $G_{17b}$ denotes the mirror image of $G_{17}$ and the blob stands for the propagator corrections from Figure 4; further,
\begin{equation}
\Sigma_2 \, = \, G_{3} \, + \, \frac{1}{2} \, G_{11} \, - \, 2 \; \, G_{12}
\, + \, G_{14} \, + \, G_{15} \, - \, 2 \; G_{16} \, + \, 4 \; G_{18} \, .
\end{equation}

We have organized the result in this form because
\begin{equation}
\Sigma_1 \, = \, O(\epsilon) \, = \, \Sigma_2 \, .
\end{equation}
These two conditions follow from the protectedness of the half-BPS
operator $\cO_{\mathbf{105}}$ (DL $[040]$). It has highest weight chiral projection $\T(\Phi^1)^4$ or, in $\cN=4$ notation, $\T(W_{12})^4$. In \cite{Penati:1999ba} it has been shown that $\langle\T(\Phi^1)^4\; \T(\bar\Phi_1)^4  \rangle_{g^4}$ contains only contact, i.e. $O(\ep)$ terms. We can assume that this is true for any other projection of $\cO_{\mathbf{105}}$. For our purposes it is most convenient to chose the projection obtained by the $SU(4)$ lowering operator $T_3{}^2$ which replaces a lower index $2$ in $W_{AB}$ by $3$, e.g., $T_3{}^2W_{12} = W_{13}$, i.e.,  $T_3{}^2 \Phi^1 = \Phi^2$. Acting with $T_3{}^2$ on $\T(\Phi^1)^4$ twice, we convert it into $2 \T(\Phi^1 \Phi^1 \Phi^2 \Phi^2) + \T(\Phi^1 \Phi^2 \Phi^1 \Phi^2)$. When evaluating the two-point function of this chiral operator, we encounter two different color structures corresponding to the sums of graphs $\Sigma_1$
and $\Sigma_2$ (the phenomenon is described in \cite{Penati:2000zv}). Thus, protectedness of $\cO_{\mathbf{105}}$ implies that $\Sigma_1=O(\ep)$ and $\Sigma_2=O(\ep)$ separately. \newline

Graphs $G_{1}$, $G_{4}$ and $G_{11}$ are products of the one-loop diagrams from Figure 1:
\begin{equation}
G_{1} \, = \, A^2 \, , \qquad G_{4} \, = \, A \, C \, , \qquad G_{11} \, =
\, C^2 \, .
\end{equation}
We have already encountered the same graphs in equation
(\ref{BBYYAC}) above:
\begin{equation}
BB_2 (YY_2 \, - \, BB_2) \  = \, 3 N (2 \, A \, + \, C) \, 3 N
(\frac{1}{3} \, C) \, = \,12 N^2 (\frac{1}{2} \, A \, C +
\frac{1}{4} \, C^2)
\end{equation}
The term $C^2$ is $O(\epsilon^2)$ and can be neglected. Even
though $A C = O(1)$ and thus cannot be dropped, it cancels against a similar term in $YY_4 - BB_4$ in the second line of eq.
(\ref{master2}):
\begin{equation}
YY_4 \, - \, BB_4 \, - \, BB_2 \,( YY_2 \, - \, BB_2) \, = \, 12
N^2 \, \left( \frac{1}{2} \, \frac{A^2}{\Pi_{12}^4} \, - \,
\frac{G_{2}}{\Pi_{12}^3} \right) \, + \, O(\epsilon) \, .
\end{equation}

The evaluation of the Grassmann integration for graph $G_{2}$ is
trivial - it actually uses twice the same trick as for diagram
A: The two parallel propagators coming in from the outer points
are evaluated at outer $\theta, \bar \theta = 0$, hence saturation
of the spinor integral at the vertices at the inner end of the
double propagators produces box operators on the subsequent lines.
We find
\begin{equation}
G_{2} \, = \, c_0^7 \, \int \frac{dx_{3,5} }{(x_{13}^4 \,  x_{23}^2 \, x_{35}^2 \,
x_{15}^2 \,  x_{52}^4)^{(1-\epsilon)}}  \, .
\end{equation}
This is a genuine four-loop integral of the $\phi^4$ type. A method for
evaluation of such integrals was developed in Dubna in the late 1970s \cite{Dubna}. The result in the $\overline{MS}$ scheme is\footnote{We are indebted to D. Kazakov for help on this point.}
\begin{equation}
G_{2} \, = \, \frac{1}{(4 \pi)^8} \, \frac{1}{4 \epsilon^3} \, \bigl(1 \,
+ \, 14 \, \epsilon \, + \, (121 - 2 \zeta(2)) \, \epsilon^2 \, + \,
O(\epsilon^3)\bigr) \, (q^2)^{(1-4 \epsilon)} \, .
\end{equation}
Like in the case of the integral $J$ we return to $x$-space by the inverse
Fourier transform and divide out three powers of the propagator:
\begin{equation}
\frac{G_{2}}{\Pi_{12}^3} \, = \, \frac{1}{4 (4 \pi^2)^2} \, \, \left(
\frac{1}{2 \epsilon^2} \, + \, \frac{5}{4 \epsilon} \, + \,
\frac{5}{4} \, + \, \frac{\zeta(2)}{2} \, + \, O(\epsilon) \right) \,
(x_{12}^2)^{(2 \epsilon)} \label{zeta2Sec}
\end{equation}
Combined with the result for graph A (\ref{evA}):
\begin{equation}
12 N^2 \, \left( \frac{1}{2} \, \frac{A^2}{\Pi_{12}^4} \, - \,
\frac{G_{2}}{\Pi_{12}^3} \right) \, = \, - \frac{3 N^2}{(4 \pi^2)^2} \,
\left( \frac{1}{4 \epsilon} \, + \, \frac{3}{4} + O(\epsilon)
\right) (x_{12}^2)^{(2 \epsilon)} \label{lastp}
\end{equation}
Finally, equation (\ref{master2}) becomes:
\begin{eqnarray}
\frac{BB_0}{YY_0} \; R & = & 1 \, + \, \frac{1}{2} \,
\frac{g^2 N}{4 \pi^2} \, + \, \frac{1}{4} \, \frac{g^4 N^2}{(4 \pi^2)^2}
\\ && - \, \frac{3 g^4 N^2}{(4 \pi^2)^2} \, \left(
\frac{1}{4 \epsilon} \, + \, \frac{3}{4} \right) (
x_{12}^2)^{(2 \epsilon)} \nonumber \\ && + \, \frac{3 g^4 N^2}{(4
\pi^2)^2} \,  \, \left( \frac{1}{4 \epsilon} \, - \, \frac{1}{4}
\right) (x_{12}^2)^\epsilon \nonumber \\ && + \, g^4 \,
8 \, f_{21} \left( \ln(x_{12}^2 ) \, + \, 2 \right) +
O(\epsilon) \nonumber
\end{eqnarray}
We see that the pole itself cancels between the second and the third line,
whereas a simple logarithm persists in the $\epsilon$-expansion. On the other
hand, conformal behavior dictates the absence of any logarithm from the ratio
$R$, because $\cK_{10}$ and $\cK_{84}$ have the same anomalous dimension.
We must therefore choose
\begin{equation}
f_{21} \, = \, \frac{3 N^2}{32 (4 \pi^2)^2} \, .
\end{equation}
The value is fully consistent with the abstract analysis in
\cite{us}: One must find $f_{21} = \frac{1}{4} \gamma_1 f_{10}$.
\newline

Our last version of eq. (\ref{master2}) is
\begin{equation}
\frac{BB_0}{YY_0} \; R \, = \, 1 \, + \, \frac{1}{2} \,
\frac{g^2 N}{4 \pi^2} \, - \, \frac{5}{4} \frac{g^4 N^2}{(4
\pi^2)^2} + O(g^6) \, ,
\end{equation}
which we match with the superspace prediction (\ref{ratio2}): the
normalizations of the two-point functions ought to differ by
\begin{equation}\label{anselmi}
\frac{\Delta (\Delta - 2)}{2 g^2 \gamma_1} \, = \, 1 + g^2
\left(\frac{\gamma_1}{2} + \frac{\gamma_2}{\gamma_1}\right) + g^4
\left( \gamma_2 + \frac{\gamma_3}{\gamma_1} \right) + O(g^6) \, .
\end{equation}
(The equation was scaled down by $2 \gamma_1$ in order to set the
lowest order equal to one.) We put in $\gamma_1 = 3 N /(4 \pi^2)$
and extract
\begin{equation}
\gamma_2 \, = \, - 3 \, \frac{N^2}{(4 \pi^2)^2} \, , \qquad
\gamma_3 \, = \, \frac{21}{4} \, \frac{N^3}{(4 \pi^2)^3} \, .
\end{equation}
The value of $\gamma_2$ has first been obtained in \cite{Bianchi:2000hn};
the value of $\gamma_3$ is in agreement with \cite{Kotikov:2004er}.

\section{The BMN operator $\cK_{\mathbf{6}}$}\label{L6}

In this section we repeat the calculation, with relatively minor changes, for the next higher BMN multiplet  $\cK_{\mathbf{6}}$ with $J=1$.  For the definition and the relevant $\cN=1$ projections we refer the reader to Section \ref{subL6}.

The tree-level correlators are
\begin{eqnarray}
BB_0 & = & 2 (N^2-1)(N^2-4) \, \Pi_{12}^4 \, ,\nonumber \\
FF_0 & = & - \frac{8}{N} (N^2-1)(N^2-4) (\partial_\mu \Pi_{12})^2 \Pi_{12}\,,
\\ YY_0 &=&  8 N (N^2-1)(N^2-4) \, \Pi_{12}^5 \, . \nonumber
\end{eqnarray}
Consequently, (recall $a = -2$)
\begin{equation}
f_{10} \, = \, \frac{N}{32 \pi^2} \, .
\end{equation}
At $O(g^2)$ we find
\begin{eqnarray}
BB_2 & = & 2 N (2 \, A \, + \, 2 \, C) \, , \\
YY_2 & = & 2 N (2 \, A \, + \, \frac{5}{2} \, C)
\end{eqnarray}
and therefore $\gamma_1 \, = \, 2 N / (4 \pi^2)$, in agreement with
\cite{Beisert:2003tq}.

For the calculation of $\langle \, B \, F^\dagger \, \rangle_{g^3}$ we
swap once again $SU(4)$ projections: The choice $B' = (\Phi^1 \Phi^1 [\bar
\Phi_2, \bar \Phi_3])$ and $F' = (\Phi^1 (\nabla^\alpha \Phi^1)(\nabla_
\alpha \Phi^1))$ is more convenient. The graphs that contribute are the same
as before,\footnote{There are a few new vanishing diagrams.} just
multiplied by an additional free line.
\begin{equation}
\langle \, F' \, {B'}^\dagger \, \rangle_{g^3} \, = \, - 4 g^3 N
(N^2-1) (N^2-4) \, \left( - \, G_{II} \, + \, G_{III} \, -
G_{IV} \right) \, .
\end{equation}
Curiously, the matter sector graph $G_{I}$ does not occur. The YM sector
diagrams sum into $2 J$ as before, so that
\begin{equation}
BF_3 \, = \, - 4 N \; J \, .
\end{equation}
In (\ref{master2}) we have the term
\begin{equation}
- f_{10} (2 \, BF_3 \, + \, a \, YY_2) \, = \, 8 N \, f_{10} \,
\left( \frac{J}{\Pi_{12}^3} \, + \, \frac{A}{\Pi_{12}^2} \right) \, + \,
O(\epsilon) \, ,
\end{equation}
where we have dropped graph C from $YY_2$. Remarkably, this is the same
combination of $J$ with $A$ as above for the Konishi multiplet. Since the
one-loop anomaly coefficient $f_{10}$ is identical (by accident?), too, we
can simply scale down equation (\ref{oncol}) by a factor 3 without doing
any further calculation:
\begin{equation}
- \, f_{10} \, ( 2 \, BF_3 \, - \, 2 \, YY_2 ) \, = \, \frac{
N^2}{4 (4 \pi^2)^2} \,  \left(\frac{1}{\epsilon} \, - \, 1 \, +
\, O(\epsilon) \right) \, (x_{12}^2)^\epsilon
\end{equation}
\newline

In the genuine $O(g^4)$ part of the calculation there are no new graphs.
We find
\begin{equation}
YY_4 \, - \, BB_4 \, = \, 4 N^2 \, \left( \frac{1}{2} \, G_{1} \, +
\, G_{4} \, + \, \frac{1}{2} \, G_{11} \, - \, G_{2} \right)
\, + \, 2 N^2 \, (\Sigma_1 \, + \, \frac{1}{4} \, G_{11}) \, +
\, 36 \, \Sigma_2
\end{equation}
The sums $\Sigma_1, \Sigma_2$ are as defined in the last section.
We have written the result in this form, because $(\Sigma_1 + 1/4 \, G_{11}) =
O(\epsilon) = \Sigma_2$ follows from the protectedness of the next higher
half-BPS multiplet $\cO_{\mathbf{196}}$. It is interesting to note that the
protectedness of this operator is equivalent to that of $\cO_{\mathbf{105}}$
since graph $G_{11}$ is a contact term. The phenomenon has been observed in
a variety of cases \cite{yascom}.

Last,
\begin{equation}
BB_2 (YY_2 \, - \, BB_2) \  = \, 2 N (2 \, A \, + \, 2 \, C) \, 2 N
(\frac{1}{2} \, C) \, = \, 4 N^2 (\, A \, C + \, C^2)
\end{equation}
and by neglecting graph $G_{11} \, = \, C^2$
\begin{equation}
YY_4 \, - \, BB_4 \, - \, BB_2 \,( YY_2 \, - \, BB_2) \, = \, 4
N^2 \, \left( \frac{1}{2} \, \frac{A^2}{\Pi_{12}^4} \, - \,
\frac{G_{2}}{\Pi_{12}^3} \right) \, + \, O(\epsilon) \, .
\end{equation}
Hence also this term can be inferred from the results of
the last section just by scaling down equation (\ref{lastp}) by a factor 3:
\begin{equation}
YY_4 \, - \, BB_4 \, - \, BB_2 \,( YY_2 \, - \, BB_2)
\, = \, - \frac{N^2}{(4 \pi^2)^2} \,
\left( \frac{1}{4 \epsilon} \, + \, \frac{3}{4} + O(\epsilon)
\right) (x_{12}^2)^{(2 \epsilon)}
\end{equation}
\newline

We substitute into (\ref{master2}):
\begin{eqnarray}
\frac{a^2}{g^2} \, \frac{BB_0}{YY_0} \; R & = & 1 \, + \, \frac{1}{4} \,
\frac{g^2 N}{4 \pi^2} \, + \, \frac{1}{16} \, \frac{g^4 N^2}{(4 \pi^2)^2}
\\ && - \, \frac{g^4 N^2}{(4 \pi^2)^2} \, \left(
\frac{1}{4 \epsilon} \, + \, \frac{3}{4} \right) (
x_{12}^2)^{(2 \epsilon)} \nonumber \\ && + \, \frac{g^4 N^2}{(4
\pi^2)^2} \,  \, \left( \frac{1}{4 \epsilon} \, - \, \frac{1}{4}
\right) (x_{12}^2)^\epsilon \nonumber \\ && + \, g^4 \,
4 \, f_{21} \left( \ln(x_{12}^2 ) \, + \, 2 \right) +
O(\epsilon) \nonumber
\end{eqnarray}
The logarithm is eliminated by setting
\begin{equation}
f_{21} \, = \, \frac{N^2}{16 (4 \pi^2)^2} \, ,
\end{equation}
which conforms with the requirement $f_{21} = \frac{1}{4} \gamma_1
f_{10}$.
\newline

Finally, (\ref{master2}) becomes
\begin{equation}
\frac{BB_0}{YY_0} \; R \, = \, 1 \, + \, \frac{1}{4} \, \frac{g^2 N}{4 \pi^2}
 \, - \, \frac{7}{16} \, \frac{g^4 N^2}{(4 \pi^2)^2} + O(g^6) \, .
\end{equation}
We equate with (\ref{anselmi}) and use $\gamma_1 = 2 N / (4 \pi^2)$ to
solve:
\begin{equation}
\gamma_2 \, = \, - \frac{3}{2} \, \frac{N^2}{(4 \pi^2)^2} \, , \qquad
\gamma_3 \, = \, \frac{17}{8} \, \frac{N^3}{(4 \pi^2)^3} \, .
\end{equation}
(The result for $\gamma_2$ was originally stated in \cite{Beisert:2003tq}.)

\section{Conclusions}

In this paper we presented a three-loop calculation of the anomalous dimensions of the first two operators of the BMN family corresponding to charges $J=0,1$. Here we summarize the results, including the lower-order ones:
\begin{eqnarray}
J=0: && \gamma_1 = \frac{3N}{4\pi^2}\,, \qquad \gamma_2 \, = \, - 3 \, \frac{N^2}{(4 \pi^2)^2} \, , \qquad
\gamma_3 \, = \, \frac{21}{4} \, \frac{N^3}{(4 \pi^2)^3} \, ; \nonumber\\
J=1: && \gamma_1 = \frac{2N}{4\pi^2}\,, \qquad \gamma_2 \, = \, - \frac{3}{2} \, \frac{N^2}{(4 \pi^2)^2} \, , \qquad
\gamma_3 \, = \, \frac{17}{8} \, \frac{N^3}{(4 \pi^2)^3} \, . \label{results}
\end{eqnarray}
Our result for $\gamma_3$, $J=0$ (Konishi) agrees with that of \cite{Kotikov:2004er}. The result for $\gamma_3$, $J=1$ is new.

As explained in the Introduction, these two values of $\gamma_3$ are sufficient to test the form of the three-loop dilatation operator proposed in \cite{Beisert:2003tq,Beisert:2003ys}. We recall that in both of these references the requirement of BMN scaling was needed to fix the operator (in \cite{Beisert:2003tq} the additional assumption of integrability was made, but  later in \cite{Beisert:2003ys} it was shown to follow from the larger symmetry taken into account). Thus, we can say that our result not only confirms the validity of the dilatation operator approach, but it also provides indirect evidence for BMN scaling at three loops.

We point out that our values (\ref{results}) are exact, i.e., we have not made use of the large $N$ (planar) approximation. Yet, they coincide with the predictions of \cite{Beisert:2003tq,Beisert:2003ys} which are supposed to apply to the planar sector only. This absence of subleading $1/N$ corrections may be explained by the relative ``shortness" (small $J$) of our operators, for which the non-planar effects may not be important.

Let us finish by a few remarks about the possible further developments. Two generalizations of our calculation could be of interest. One of them would be to go to four loops. An obvious difficulty is the exponential growth of the combinatorics. However, our present experience shows that in the end these massive calculations always produce rather simple results, so one might hope this is not just a ``low loop effect". Another complication would be the occurrence of new, five- or maybe six-loop scalar integrals. In this paper we had to deal with a single difficult four-loop integral. Hopefully, the method of \cite{Dubna} is applicable to higher loops as well. Finally, at four loops we may reach the limitations of the supersymmetric dimensional reduction scheme (see \cite{Siegel:1980qs}). This scheme, although not indispensable, is highly efficient due to the manifest $\cN=1$ supersymmetry.

The other generalization would be to look at the next value $J=2$. This would be of great interest, since our results so far helped to unambiguously fix the form of the three-loop dilatation operator, but new results for $J=2$ would provide a check on a true prediction. The obvious major complication in this direction is the possibility of mixing among the various realizations of the $J=2$ BMN operator. Still, we believe that trying is worthwhile.

Finally, we should mention another interesting line of research. Instead of going in the direction of higher values of $J$ (higher twist), one might stay with twist two operators, but increase their spin (see, e.g., \cite{Beisert:2004di} for a recent discussion). Three-loop results of this type are already available \cite{Kotikov:2004er}, but our method would provide an alternative check using more conventional field theory methods. Yet another approach to higher spins consists in calculating the four-point function of the stress-tensor multiplet and extracting anomalous dimensions by OPE techniques. We believe that a generalization of the harmonic superspace approach to four-point functions developed in \cite{Eden:2000mv} could still be applicable at order $O(g^6)$.

\section*{Acknowledgments}  Most of this work was done while BE was a member
of the theory group of the Dipartimento di Fisica, Universit\`a di Roma ``Tor
Vergata", where he held a DFG postdoctoral fellowship. ES is grateful to this
group and especially to Augusto Sagnotti for extending to him their warm
hospitality at ``Tor Vergata". We are deeply indebted to Yassen Stanev who
participated at the initial stages of this project and has continuously
shared his experience with us. We also profited from numerous discussions
with G. Arutyunov, N. Beisert, M. Bianchi, T. Binoth, D. Kazakov, G.C. Rossi,
Ch. Schubert, M. Staudacher, F. Tkachov. This work was supported in part by
the MIUR-COFIN contract 2003-023852 and by the INTAS contract 00-00254.

\newpage

\section{Appendix}

\subsection{Feynman rules and conventions}\label{3.1}

The classical ${\cN=4}$ SYM action formulated in terms of $\cN=1$ superfields
has the form
\begin{eqnarray}
  S_{\cN=4}&=& \int d^4x \, d^2\theta \, d^2\bar\theta\; \T \left(e^{gV} \bar \Phi_I e^{-gV} \Phi^I  \right) \label{67}  \\
  &+&  \frac{1}{4}\int d^4x_L \, d^2\theta\; \T(W^\alpha W_\alpha) + \left[
\frac{g}{3!}\int d^4x_L \, d^2\theta\; \ep_{IJK} \T (\P^I[\P^J,\P^K]) + c.c. \right]\,. \nonumber
\end{eqnarray}
Here
\begin{equation}\label{68}
  x^\mu_L = x^\mu +  i\q\sigma^\mu\bq\,, \qquad x^\mu_R = x^\mu - i\q\sigma^\mu\bq
\end{equation}
are the space-time coordinate in the (left-handed) chiral and (right-handed)
antichiral bases of superspace. All the superfields are in the adjoint
representation of the gauge group $SU(N)$, and the generators and the
structure constants satisfy the relations
\begin{equation}\label{69}
\T ( T^a T^b ) = \delta^{ab}\,, \qquad f^{abc}f^{abd} = 2 N \, \delta^{cd}\, .
\end{equation}
To this classical action one should add (Feynman) gauge fixing and ghost
terms. For our purposes we only need the cubic vertices (except the $VVV$ one).
 The relevant part of the gauge-fixed action is
\begin{eqnarray}
  S^{g.f.}_{\cN=4}&=& \int d^4x \, d^2\theta \, d^2\bar\theta\; \T \Bigl[\bar \Phi_I \Phi^I - \frac{1}{2} V\square V + g \, [V,\bP_I]\P^I  \Bigr]  \label{70}  \\ && + g\int d^4x_L \, d^2\theta\; \T (\P^1[\P^2,\P^3])  - g \int d^4x_R \, d^2\bq\; \T (\bP_1[\bP_2,\bP_3]) + \ldots \,. \nonumber
\end{eqnarray}

Our spinor conventions are \cite{harmonic}:
\begin{equation}
\psi^\a \, = \, \epsilon^{\a\b} \psi_\b, \qquad \psi_\a \, = \, \epsilon_{\a\b} \psi^\b, \qquad \epsilon_{12} \, = \, 1, \qquad \epsilon_{\a\b} \epsilon^{\b\gamma} \, = \, \delta_\a^\gamma
\end{equation}
and exactly the same with dotted indices. Complex conjugation replaces an
undotted by a dotted index and vice versa; however, it does not exchange
lower and upper position.

The $2\times 2$ sigma matrices are Hermitian. The $\tilde \sigma$ matrix is
obtained from $\sigma_{\a \dot \a}$ by raising of both indices as defined by
the last equation. They satisfy the following relations:
\begin{eqnarray}
& \sigma^\mu \tilde \sigma^\nu \, = \, \eta^{\mu\nu} - i \sigma^{\mu\nu} \, ,  &
  \tilde \sigma^\mu \sigma^\nu \, = \, \eta^{\mu\nu} - i \tilde \sigma^{\mu\nu} \, ,\\
& (\sigma^\mu)_{\a \dot \a} (\tilde \sigma_\mu)^{\dot \b \b} \, = \, 2
\delta_\a^\b \delta_{\dot \a}^{\dot \b} \, ,  & (\sigma^\mu)_{\a \dot \a}
(\tilde \sigma_\nu)^{\dot \a \a} \, = \, 2 \delta^\mu_\nu \nonumber \, .
\end{eqnarray}

In $\cN = 4$ supersymmetry, spinors carry $SU(4)$ indices with canonical positions $\q_A, \, \bq^A$. Complex conjugation makes a
lower internal index go up and an upper go down, while the sequence of
indices of a tensor is not changed. Last, $\epsilon_{1234} =
\epsilon^{1234} = 1$.

The superspace covariant derivatives are
\begin{equation}
D_\a \, = \, \partial_\a + i \bq^{\dot \a} \partial_{\a \dot \a}, \qquad
\bar D_{\dot \a} \, = \, - \partial_{\dot \a} - i \q^\alpha \partial_{\a
\dot \a}, \qquad \partial_{\a\dot\a} = \partial_\mu (\sigma^\mu)_{\a\dot\a}  \,.
\end{equation}
The derivatives with upper indices are obtained by raising in the usual
fashion. A useful shorthand is
\begin{equation}
\q^2 \, = \, \q^\a \q_\a, \qquad \bq^2 \, = \, \bq_{\dot \a} \bq^{\dot \a},
\end{equation}
and similarly for the product of two different spinors. Consequently,
\begin{equation}
D \cdot D \; \q^2 \, \equiv \, - \frac{1}{4} D^2 \,
\q^2 \, = \, 1, \qquad \bar D \cdot \bar D \; \bq^2 \, \equiv
\, - \frac{1}{4} \bar D^2 \, \bq^2 \,
= \, 1.
\end{equation}
Under the $x$-integral we may thus identify
\begin{equation}
d^2\q \, = \, D \cdot D, \qquad d^2\bq \, = \, \bar D \cdot \bar D \, .
\end{equation}

We are going to perform our graph calculations in $x$ space,\footnote{Occasionally, when we need to evaluate some complicated integrals, we switch over to momentum space where we can use the powerful {\it Mincer} algorithm.} since the supersymmetry algebra is easier there. The propagator of the chiral superfields is
\begin{equation}
\langle \Phi^I(1) \bar \Phi_J(2) \rangle \, = \, \delta^I_J \, \Pi_{12} \, ,
\qquad \Pi_{12} \, = \, - \frac{1}{4 \pi^2 \hat x_{12}^2} \label{71} \, .
\end{equation}
(We find it more convenient not to write the gauge group trace $\delta^{ab}$
--- it is there, of course.) In the supersymmetric dimensional reduction (SSDR)
scheme with $D = 4 - 2 \epsilon$, the last formula becomes
\begin{equation}
\Pi_{12} \, = \, - \frac{\Gamma(1-\epsilon) \pi^\epsilon}{4 \pi^2 (\hat x_{12}^2)^{(1 - \epsilon)}} \, .
\end{equation}
In these expressions we use the coordinate difference
\begin{equation}\label{72}
  \hat x^\mu_{12} = x^\mu_{1L} - x^\mu_{2R} - 2i\q_1\sigma^\mu\bq_2 \, ,
\end{equation}
which is manifestly chiral/antichiral and invariant under the supersymmetry
transformations
\begin{equation}
\delta \, x \, = \, - i \, \epsilon \sigma \bq + i \, \q \sigma \bar
\epsilon \, , \qquad \delta \q \, = \, \epsilon \, , \qquad \delta \bq \, =
\, \bar \epsilon \, .
\end{equation}
The various nilpotent shifts in the propagator (\ref{71}) can conveniently be
assembled into an exponential factor:
\begin{equation}\label{721}
\frac{1}{\hat x_{12}^2} \, = \, \exp\left\{i \,[\theta_1 \sigma_\mu \bar
\theta_1  +  \theta_2 \sigma_\mu \bar \theta_2  - 2  \theta_1
\sigma_\mu  \bar \theta_2] \, \partial^\mu_1\right\} \frac{1}{x_{12}^2} \,  \equiv \, e^{\Delta_{12}} \frac{1}{x_{12}^2}
\end{equation}
Our technique for evaluating supergraphs is to expand these exponential
shifts in $\q, \bq$ and to integrate out the Grassmann variables. What is left
is a multiple integral of some differential operator acting on a product
of $x$-space propagators. Simplifications within a sum of graphs can usually
be demonstrated just by partial integration in configuration space.
The matter propagator satisfies the Green's function property
\begin{equation}
\square \, \Pi_{12} \, = \, \delta^D(\hat x_{12}) \label{green}
\end{equation}
which can often be used to remove an integration.

Correlation functions in Minkowski space have the weight factor $e^{i S}$
under the path integral. This would usually mean to put a factor $i$ on every
vertex in the action (\ref{70}). On the other hand, the $i$ is clearly
absorbed into the integration measure when Wick-rotating to Euclidean
space. It is therefore a consistent prescription to take the vertices from
(\ref{70}) without introducing extra $i$'s and to simply evaluate all
configuration space integrals in Euclidean signature. Note that the Green's
function property (\ref{green}) is metric independent.
\newline

The propagator for the gauge superfield in Feynman gauge is
\begin{equation}\label{73}
\langle V(1) V(2)\rangle = + \frac{\theta_{12}^2 \bar \theta_{12}^2}{4 \pi^2
x_{12}^2} \, .
\end{equation}
(We omitted $\delta^{ab}$ once again.)
Due to the Grassmann delta functions in the numerator, we could also use
$(x_{1L} - x_{2L})^2$ or $(x_{1R}-x_{2R})^2$ in the denominator.

The regularized version of (\ref{73}) picks up a factor $\Gamma(1-\epsilon)
\pi^\epsilon$, and $1/x^2$ is modified to $1/(x^2)^{(1-\epsilon)}$. The
numerator is not changed - the very essence of the SSDR scheme
is to handle the spinors and the Dirac algebra like in four dimensions.

The graphs $G_I$ and $G_{II}$ from Section \ref{g3s} contain the structure
\begin{eqnarray}
(\partial_i \partial_j \partial_k \partial_l) &\equiv& \T(\sigma^\mu\tilde\sigma^\nu\sigma^\lambda\tilde\sigma^\rho)\;  \pa_{i\mu}\pa_{j\nu}\pa_{k\lambda}\pa_{l\rho} \label{723}\\
  &=& 2[(\pa_i \cdot \pa_j)(\pa_k \cdot \pa_l) - (\pa_i \cdot \pa_k)(\pa_j \cdot \pa_l) + (\pa_i \cdot \pa_l)(\pa_j \cdot \pa_k) ] \nonumber\\
  && -2i \ep_{\mu\nu\lambda\rho}\pa_i^\mu\pa_j^\nu\pa_k^\lambda\pa_l^\rho \,.\nonumber
\end{eqnarray}
with $(\pa_i \cdot \pa_j) \, = \, \pa_i^\mu \partial_{j\mu}$. The four-index
antisymmetric tensor drops in the integral $J$ since it is a pseudo-tensor: the
final result for this two-point integral can only depend on $x^\mu_{12}$
(or the outer momentum, respectively) and out of one vector one cannot
construct a pseudo-scalar. The other graphs we actually have to evaluate do not
involve any complicated trace structures. SSDR may become inconsistent when a contraction of two antisymmetric tensors occurs \cite{Siegel:1980qs}.
Our calculation stays clear of this problem.
\newline

To give an impression of the formalism we evaluate the one-loop diagrams A
and C from Figure 1. Graph A is particularly simple:
\begin{equation}
A \, = \, \int d^Dx_{R3} \, d^2\bq_3 \, d^Dx_{L4} \, d^2\q_4 \, \Pi_{13}^2 \,
\Pi_{43} \, \Pi_{42}^2
\end{equation}
We want the $\q_1,\bq_1, \q_2, \bq_2 \, = \, 0$ component of the
supergraph, i.e. an exchange between the physical scalars at the outer points.
Let us inspect the nilpotent shifts of the various propagators. We find
\begin{equation}
\hat x_{13} \, = \, x_1 - x_{3R} \, , \qquad \hat x_{42} = x_{4L} - x_2
\end{equation}
exactly because the outer thetas are zero. Hence we do not have to expand
these propagators. The middle line has the shift
\begin{equation}
\hat x_{43}\, = \, x_{4L} - x_{3R} - 2 i \q_4 \sigma \bq_3
\end{equation}
which can be written in the exponential form
\begin{equation}
e^{-2 i (\q_4 \partial_4 \bq_3)} \frac{1}{(x_{4L3R}^2)^{(1-\epsilon)}} \, = \,
- \q_4^2 \, \bq_3^2 \, \square_4 \frac{1}{(x_{4L3R}^2)^{(1-\epsilon)}} =
- \frac{\q_4^2 \, \bq_3^2}{c_0} \, \delta^D(x_{4L3R}) \, .
\end{equation}
Here $c_0 = - \Gamma(1-\epsilon) \pi^\epsilon / (4 \pi^2)$ comes from the
standard scalar propagator. In the $x$-space $\overline{MS}$ scheme (\ref{msx})
we obtain:
\begin{eqnarray} \label{evA}
A & = & - \, c_0^4 \, \int \frac{d^{D}x_3}{(x^2_{13}x^2_{23})^{2(1-\ep)}} \\ & = &  - \frac{1}{2 (4 \pi^2)}
\, \left( \frac{1}{\epsilon} \, + \, 1 \, + \, \frac{3 \zeta(2)}{2} \,
\epsilon \, +
\, O(\epsilon^2) \right) \, \frac{1}{(4 \pi^2 x_{12}^2)^2} \, (x_{12}^2)^{(3
\epsilon)} \nonumber
\end{eqnarray}
Next, we consider graph C:
\begin{equation}
\int d^Dx_{5,6} \, d^2\q_{5,6}\, d^2\bar\q_{5,6} \, \Pi_{15} \, \Pi_{52} \, \Pi_{16} \, \Pi_{62} \,
\langle V(5) V(6) \rangle
\end{equation}
We simplify the calculation by a trick: let us shift the integration variables
$x_5,x_6$ to $x_{5R},x_{6R}$. The Jacobian for such a nilpotent shift is 1.
Since once again $\q_1,\bq_1, \q_2, \bq_2 \, = \, 0$, the propagators
connecting to point 1 will no longer be expanded; there is also no
derivative acting on the YM line. For the other two lines we find
\begin{equation}
\Bigl(e^{- 2 i (\q_5 \partial_2 \bq_5)} \frac{1}{x_{25}^2}\Bigr) \Bigl(e^{- 2 i
(\q_6 \partial_2 \bq_6)} \frac{1}{x_{26}^2}\Bigr)
\end{equation}
The numerator of the YM propagator behaves like a $\q$ delta-function. We can
therefore replace $\q_6 \rightarrow \q_5, \bq_6 \rightarrow \bq_5$ in the last
equation. The thetas square up and yield a trace on the derivatives which is
a perfect square. It follows,
\begin{equation}
C \, = \, + c_0^5 \, \square_2 \, \int \frac{d^Dx_5 \, d^Dx_6}{(x_{15}^2
x_{25}^2 x_{56}^2 x_{16}^2 x_{26}^2)^{(1-\epsilon)}}
\end{equation}
where one minus comes from the sign of the YM propagator (we have often
absorbed this into the combinatorics) and the second one from the $i^2$.
The integral itself is in fact finite ($O(\epsilon^0)$ and
higher). In dimensional regularization its functional form is
$1/(x_{12}^2)^{(1-3\epsilon)}$. We conclude
\begin{equation}
C \, = \, O(\epsilon) \frac{1}{(x_{12}^2)^{(2-3\epsilon)}} \, .
\end{equation}

\subsection{Decoupling of the BPS operators ${\cal
D}_{84}$ and ${\cal D}_{300}$} \label{AppB}

We first discuss the $J=0$ case. The bare second level descendant
of the Konishi operator is (\ref{15})
\begin{equation}
{\cal Y} \, = \, \T([\Phi^1, \Phi^2][\Phi^1, \Phi^2]) \nonumber \, ,
\end{equation}
which is tree-level and one-loop orthogonal to
\begin{equation}
{\cal D}_{84} \, = \, \T(\Phi^1 \Phi^1)\T(\Phi^2 \Phi^2) -
\T(\Phi^1 \Phi^2)\T(\Phi^1 \Phi^2) \, - \, \frac{1}{N} \, {\cal
Y} \, ,
\end{equation}
c.f. \cite{Bianchi:1999ge}. The operator ${\cal D}_{84}$ is one-loop protected
since in its two-point function graph A drops out and only
graph C survives.

We have to show that at $O(g^4)$, too,
${\cal D}_{84}$ is a pure state and that it stays orthogonal to ${\cal
Y}$; implying that ${\cal Y}$ remains an eigenvector up to $O(g^4)$
without receiving any admixtures of ${\cal D}_{84}$.

Calculation of the combinatorics yields
\begin{eqnarray}
\langle \, {\cal D}_{84} \, {\cal Y} \, \rangle_{g^4} & = & 72 N
\, (N^2-1)(N^2-4) \, \Sigma_2 \, , \\ \langle \, {\cal D}_{84} \,
{\cal D}_{84} \, \rangle_{g^4} & = & 24 N^2 (N^2-1)(N^2-4) \, (2
N^2 \, \Sigma_1 \, + \, (N^2-3) \, \Sigma_2) \, , \nonumber
\end{eqnarray}
with the sums of graphs $\Sigma_1, \Sigma_2$ from Section
(\ref{334}). Thereby, both correlators vanish up to contact terms.

For our other example with $J=1$ the situation is strictly
analogous: There is the second level descendant (\ref{y300})
\begin{equation}
{\cal Y} \, = \, - 2 \T(\Phi^1 \Phi^1 [\Phi^1, \Phi^2] \Phi^2)
\nonumber
\end{equation}
of the operator ${\cal L}_6$ and the 1/4 BPS operator
\begin{equation}
{\cal D}_{300} \, = \, \T(\Phi^1 \Phi^1 \Phi^1) \T(\Phi^2
\Phi^2) - 2 \T(\Phi^1 \Phi^1 \Phi^2) \T(\Phi^1 \Phi^2) + \T(\Phi^1
\Phi^2 \Phi^2) \T(\Phi^1 \Phi^1) \, + \, \frac{3}{N} \, {\cal
Y} \nonumber
\end{equation}
which is tree- and one-loop orthogonal to ${\cal Y}$ \cite{Ryzhov:2001bp}.

At two-loop level we have
\begin{eqnarray}
\langle \, {\cal D}_{300} \, {\cal Y} \, \rangle_{g^4} & = & - 96
\, (N^2-1)(N^2-4)(N^2-9) \, \Sigma_2 \, , \\ \langle \, {\cal
D}_{300} \, {\cal D}_{300} \, \rangle_{g^4} & = & \frac{12
(N^2-1)(N^2-4)(N^2-9)}{N} \nonumber \\ && * \, (10 N^2 \, (\Sigma_1
+ \frac{1}{4} \, G_{11}) \, + \, (2 N^2 - 24) \, \Sigma_2) \, .
\nonumber
\end{eqnarray}
thus proving our assertions. Remarkably, the sum
$\Sigma_1$ is shifted by $1/4 \, G_{11} = 1/4 \, C^2 =
O(\epsilon^2)$ exactly as in the two-loop two-point function of
the 1/2 BPS single trace state $O_{196}$, see Section \ref{L6}.

\subsection{The mass scale $\mu$ and the $\overline{MS}$ scheme} \label{AppC}

We work in $D = 4 - 2 \epsilon$ dimensions. Recall the configuration space
propagator (\ref{propdimreg})
\begin{equation}
\Pi_{12} \, = \, - \frac{\Gamma(1-\epsilon) \, \pi^\epsilon
(\mu^2)^\epsilon}{4 \pi^2 (x_{12}^2)^{(1-\epsilon)}} \nonumber
\end{equation}
into which we have introduced the mass scale $\mu$ in order to keep the canonical
dimension of the chiral superfield $\Delta = 1$. Note that
\begin{equation}
\Gamma(1-\epsilon) \, = \, e^{\gamma \epsilon} \bigl(1+\epsilon^2
\frac{\zeta(2)}{2} + \epsilon^3 \frac{\zeta(3)}{3} + \ldots \bigr)
\, ,
\end{equation}
so that we can write
\begin{equation}
\Pi_{12} \, = \, - \frac{1}{4 \pi^2 x_{12}^2} \, (\bar \mu^2
x_{12}^2)^\epsilon \, \bigl(1 + \epsilon^2 \frac{\zeta(2)}{2} +
O(\epsilon^3)\bigr)
\end{equation}
which defines the $\overline{MS}$ scheme mass scale
\begin{equation}
\bar \mu^2 \, = \, \mu^2 \, \pi e^\gamma \, . \label{msx}
\end{equation}
In this scheme the fractional powers of $\pi$ and the
Euler-Mascheroni constant $\gamma$ are made to disappear from the
results for Feyman integrals. We use this definition throughout the paper.
Nevertheless, one could do slightly better: the form of the propagator
actually suggests to define
\begin{equation}
(\tilde \mu^2)^\epsilon \,= \, (\mu^2 \pi)^\epsilon
\Gamma(1-\epsilon) \, .
\end{equation}
This re-definition of the mass scale additionally makes the
$\zeta(2) = \pi^2/6$ disappear from our formulas (\ref{sudden}),
(\ref{zeta2Sec}), (\ref{evA}).
\newline

The $\overline{MS}$ scheme defined in (\ref{msx}) is particular to
configuration space; it is inequivalent to the usual definition in
momentum space, where
\begin{equation}
\bar \mu^2 \, = \, \mu^2 \, 4 \pi e^{-\gamma}
\end{equation}
helps to eliminate the Euler-Mascheroni constant as well as
fractional powers of 4 and $\pi$ from the results. It was pointed
out to us by D. Kazakov that the choice
\begin{equation}
(\tilde \mu^2)^\epsilon \, = \, \frac{(\mu^2 \, 4
\pi)^\epsilon}{\Gamma(1-\epsilon)}
\end{equation}
also hides all $\zeta(2)$, which inspired our second $x$-space
prescription above. \newline

As we have kept the dimension of the fields at their canonical values,
every vertex carries a factor $(\mu^2)^{-\epsilon}$ in $x$-space (or its
inverse in momentum space). In order to simplify the
notation we have suppressed the mass scale
throughout the paper. It can simply be restored after integration by
completing any fractional power like
\begin{equation}
(x^2)^\epsilon \, \rightarrow \, (\mu^2 x^2)^\epsilon \, , \qquad
(q^2)^\epsilon \, \rightarrow \,
\bigl(\frac{q^2}{\mu^2}\bigr)^\epsilon \, . \label{rinst}
\end{equation}

Let us illustrate the change of schemes on the example of the
integral from graph A:
\begin{equation}
h(x_{12}) \, = \, c_0^4 \, \int \frac{d^Dx_3}{(x_{13}^2
x_{23}^2)^{2 (1-\epsilon)}}
\end{equation}
The integral can be evaluated directly in $x$-space, e.g. by the
standard Feynman parameter trick. We rather choose to mimic the
steps needed to use \emph{Mincer}; therefore we go through
momentum space and then transform back. The forward Fourier-transform is
\begin{equation}
\int d^Dx \, e^{i q x} \, \frac{1}{(x^2)^\alpha} \, = \,
4^{\frac{D}{2}-\alpha} \, \pi^{\frac{D}{2}} \,
\frac{\Gamma(\frac{D}{2}-\alpha)}{\Gamma(\alpha)} \,
\frac{1}{(q^2)^{(\frac{D}{2} - \alpha)}} \, .
\end{equation}
The backward Fourier-transform has the same form with the roles of $x$ and $q$
exchanged, but it comes with a factor $1/(2 \pi)^D$. In our integral $h$
we replace every $x$-space propagator by its back transform from momentum
space and then integrate out $x_3$:
\begin{equation}
h(x_{12}) \, = \, \int \frac{(\mu^2)^\epsilon d^Dq}{(2 \pi)^D} \,
e^{i q x_{12}} \, H(q) \, , \qquad H(q) \, = \, \Bigl[ \int
\frac{(\mu^2)^\epsilon d^Dp}{(2 \pi)^D \, p^2 (p-q)^2} \Bigr]^2 \, .
\end{equation}
We have reinstated the mass scale on the measure
since we intend to demonstrate explicitly how redefinitions of $\mu^2$ help to
cast the result into a tidier form. The subintegral in $H(q)$ is again
elementary. We find:
\begin{eqnarray}
H(q) & = & \frac{1}{(4 \pi)^4} \Bigl(\frac{q^2}{4 \pi
\mu^2}\Bigr)^{(- 2 \epsilon)} \frac{\Gamma(\epsilon)^2
\Gamma(1-\epsilon)^4}{\Gamma(2-2\epsilon)^2} \\ & = & \frac{1}{(4
\pi)^4} \Bigl(\frac{q^2}{4 \pi \mu^2}\Bigr)^{(- 2 \epsilon)} \,
\bigl( \frac{1}{\epsilon^2} +  \frac{4 - 2 \gamma}{\epsilon} + (12
- 8 \gamma + 2 \gamma^2 - \zeta(2)) + O(\epsilon) \bigr) \nonumber
\\ & = & \frac{1}{(4 \pi)^4} \Bigl(\frac{q^2}{\bar \mu^2}\Bigr)^{
(- 2 \epsilon)} \, \bigl( \frac{1}{\epsilon^2} +
\frac{4}{\epsilon} + (12 - \zeta(2)) + O(\epsilon) \bigr)
\nonumber \\ & = & \frac{1}{(4 \pi)^4} \Bigl(\frac{q^2}{
\tilde \mu^2}\Bigr)^{ (- 2 \epsilon)} \, \bigl(
\frac{1}{\epsilon^2} + \frac{4}{\epsilon} + 12 + O(\epsilon)
\bigr) \nonumber
\end{eqnarray}
In the second to the last line the momentum space $\overline{MS}$
mass scale is used, in the last line the further modified
$\tilde{}$ version of it.

The $\overline{MS}$ result could also be taken from \emph{Mincer}:
the integral can e.g. be obtained by topology \textbf{T1} with $p_5^2$ in
the numerator. In general, the program's results are to be multiplied
with a factor $1/(4 \pi)^{2 l}$, where $l$ is the loop order as seen
in momentum space. Next, we undo the $p$-space $\overline{MS}$ scheme.
We should remark, however, that the reconstruction of the $\zeta(2)$ terms
is not obvious. Luckily, these are $O(\epsilon^2)$ subleading, and
thus the part of the integral $J$ needed in our calculations is
insensitive to the issue.

Finally, we transform back to configuration space, where we change to the new
$\overline{MS}$ scheme (\ref{msx}) or, if desired, to its modified version
with $\tilde \mu^2$.
\begin{eqnarray}
h(x_{12}) & = & \frac{1}{(4 \pi^2)^3} \, \frac{(\pi \mu^2
x_{12}^2)^{(3 \epsilon)}}{(x_{12}^2)^2} \, \frac{\Gamma(2-3
\epsilon) \Gamma(\epsilon^2) \Gamma(1-\epsilon)^4}{4 \, \Gamma(2
\epsilon) \Gamma(2 - 2\epsilon)^2} \\ & = & \frac{1}{(4 \pi^2)^3}
\, \frac{(\pi \mu^2 x_{12}^2)^{(3 \epsilon)}}{(x_{12}^2)^2} \,
\bigl(\frac{1}{2\epsilon} + \frac{1 + 3 \gamma}{2} + \frac{12
\gamma + 9 \gamma^2  + 3 \zeta(2)}{4} \, \epsilon + O(\epsilon^2)
\bigr) \nonumber \\ & = & \frac{1}{(4 \pi^2)^3} \, \frac{(\bar
\mu^2 x_{12}^2)^{(3 \epsilon)}}{(x_{12}^2)^2} \,
\bigl(\frac{1}{2\epsilon} + \frac{1}{2} + \frac{3 \zeta(2)}{4} \,
\epsilon + O(\epsilon^2) \bigr) \nonumber
\\ & = & \frac{1}{(4 \pi^2)^3} \, \frac{(\tilde
\mu^2 x_{12}^2)^{(3 \epsilon)}}{(x_{12}^2)^2} \,
\bigl(\frac{1}{2\epsilon} + \frac{1}{2} + 0 \, \epsilon +
O(\epsilon^2) \bigr) \nonumber
\end{eqnarray}
Last, we remark that these rescalings of $\mu^2$ cannot influence
our results for the anomalous dimensions, because the logarithms
--- and with it the mass scale --- drop from the defining equation
(\ref{master2}). In fact, in ${\cal N}=4$ the anomalous dimensions
should be fully scheme independent.


\newpage

\begin{thebibliography}{99}

\bibitem{Berenstein:2002jq}
D.~Berenstein, J.~M.~Maldacena and H.~Nastase,
``Strings in flat space and pp waves from N = 4 super Yang Mills,''
JHEP {\bf 0204} (2002) 013
[arXiv:hep-th/0202021].

\bibitem{Minahan:2002ve}
J.~A.~Minahan and K.~Zarembo,
``The Bethe-ansatz for N = 4 super Yang-Mills,''
JHEP {\bf 0303} (2003) 013
[arXiv:hep-th/0212208];\\
A.~V.~Belitsky, S.~E.~Derkachov, G.~P.~Korchemsky and A.~N.~Manashov,
``Superconformal operators in N = 4 super-Yang-Mills theory,''
[arXiv:hep-th/0311104].

\bibitem{Beisert:2003tq}
N.~Beisert, C.~Kristjansen and M.~Staudacher,
``The dilatation operator of N = 4 super Yang-Mills theory,''
Nucl.\ Phys.\ B {\bf 664} (2003) 131
[arXiv:hep-th/0303060].

\bibitem{Beisert:2003yb}
N.~Beisert,
``The complete one-loop dilatation operator of N = 4 super Yang-Mills theory,''
Nucl.\ Phys.\ B {\bf 676} (2004) 3
[arXiv:hep-th/0307015];\\
N.~Beisert and M.~Staudacher,
``The N = 4 SYM integrable super spin chain,''
Nucl.\ Phys.\ B {\bf 670} (2003) 439
[arXiv:hep-th/0307042].

\bibitem{Frolov:2002av}
S.~Frolov and A.~A.~Tseytlin,
``Multi-spin string solutions in AdS(5) x S**5,''
Nucl.\ Phys.\ B {\bf 668} (2003) 77
[arXiv:hep-th/0304255];\\
G.~Arutyunov, S.~Frolov, J.~Russo and A.~A.~Tseytlin,
``Spinning strings in AdS(5) x S**5 and integrable systems,''
Nucl.\ Phys.\ B {\bf 671} (2003) 3
[arXiv:hep-th/0307191];\\
G.~Arutyunov, J.~Russo and A.~A.~Tseytlin,
``Spinning strings in AdS(5) x S**5: New integrable system relations,''
Phys.\ Rev.\ D {\bf 69} (2004) 086009
[arXiv:hep-th/0311004].

\bibitem{Beisert:2003xu}
N.~Beisert, J.~A.~Minahan, M.~Staudacher and K.~Zarembo,
``Stringing spins and spinning strings,''
JHEP {\bf 0309} (2003) 010
[arXiv:hep-th/0306139].

\bibitem{Arutyunov:2004xy}
G.~Arutyunov and M.~Staudacher,
``Matching higher conserved charges for strings and spins,''
JHEP {\bf 0403} (2004) 004
[arXiv:hep-th/0310182];\\
V.~A.~Kazakov, A.~Marshakov, J.~A.~Minahan and K.~Zarembo,
``Classical / quantum integrability in AdS/CFT,''
JHEP {\bf 0405} (2004) 024
[arXiv:hep-th/0402207];\\
G.~Arutyunov and M.~Staudacher,
``Two-loop commuting charges and the string / gauge duality,''
[arXiv:hep-th/0403077];\\
G.~Arutyunov, S.~Frolov and M.~Staudacher,
``Bethe ansatz for quantum strings,''
[arXiv:hep-th/0406256].

\bibitem{Beisert:2002ff}
N.~Beisert, C.~Kristjansen, J.~Plefka and M.~Staudacher,
``BMN gauge theory as a quantum mechanical system,''
Phys.\ Lett.\ B {\bf 558} (2003) 229
[arXiv:hep-th/0212269].

\bibitem{Beisert:2003ys}
N.~Beisert,
``The su(2$|$3) dynamic spin chain,''
Nucl.\ Phys.\ B {\bf 682} (2004) 487
[arXiv:hep-th/0310252].

\bibitem{Gross:2002su}
D.~J.~Gross, A.~Mikhailov and R.~Roiban,
``Operators with large R charge in N = 4 Yang-Mills theory,''
Annals Phys.\  {\bf 301} (2002) 31
[arXiv:hep-th/0205066].

\bibitem{Santambrogio:2002sb}
A.~Santambrogio and D.~Zanon,
``Exact anomalous dimensions of N = 4 Yang-Mills operators with large R
charge,''
Phys.\ Lett.\ B {\bf 545} (2002) 425
[arXiv:hep-th/0206079].

\bibitem{Klose:2003qc}
T.~Klose and J.~Plefka,
``On the integrability of large N plane-wave matrix theory,''
Nucl.\ Phys.\ B {\bf 679} (2004) 127
[arXiv:hep-th/0310232].

\bibitem{Vogt:2004mw}
A.~Vogt, S.~Moch and J.~A.~M.~Vermaseren,
``The three-loop splitting functions in QCD: The non-singlet case,''
Nucl.\ Phys.\ B {\bf 688} (2004) 101
[arXiv:hep-ph/0403192].

\bibitem{Kotikov:2004er}
A.~V.~Kotikov, L.~N.~Lipatov, A.~I.~Onishchenko and V.~N.~Velizhanin,
``Three-loop universal anomalous dimension of the Wilson operators in N = 4
SUSY Yang-Mills model,''
Phys.\ Lett.\ B {\bf 595} (2004) 521
[arXiv:hep-th/0404092].

\bibitem{Anselmi:1998ms}
D.~Anselmi,
``The N = 4 quantum conformal algebra,''
Nucl.\ Phys.\ B {\bf 541} (1999) 369
[arXiv:hep-th/9809192].

\bibitem{Eden:2003sj}
B.~Eden,
``On two fermion BMN operators,''
Nucl.\ Phys.\ B {\bf 681} (2004) 195
[arXiv:hep-th/0307081].

\bibitem{Intriligator:1999ff}
K.~A.~Intriligator and W.~Skiba,
``Bonus symmetry and the operator product expansion of N = 4
super-Yang-Mills,''
Nucl.\ Phys.\ B {\bf 559} (1999) 165
[arXiv:hep-th/9905020].

\bibitem{Bianchi:2001cm}
M.~Bianchi, S.~Kovacs, G.~Rossi and Y.~S.~Stanev,
``Properties of the Konishi multiplet in N = 4 SYM theory,''
JHEP {\bf 0105} (2001) 042
[arXiv:hep-th/0104016].

\bibitem{Clark:1978jx}
T.~E.~Clark, O.~Piguet and K.~Sibold,
``Supercurrents, Renormalization And Anomalies,''
Nucl.\ Phys.\ B {\bf 143} (1978) 445;\\
K.~Konishi,
``Anomalous Supersymmetry Transformation Of Some Composite Operators In Sqcd,''
Phys.\ Lett.\ B {\bf 135} (1984) 439.

\bibitem{us}
B.~Eden, C.~Jarczak, E.~Sokatchev and Y.~Stanev,
``Operator mixing at level $g^4$ in $\cN=4$ SYM: The Konishi anomaly
revisited", in preparation.

\bibitem{Bianchi:1999ge}
M.~Bianchi, S.~Kovacs, G.~Rossi and Y.~S.~Stanev,
``On the logarithmic behavior in N = 4 SYM theory,''
JHEP {\bf 9908} (1999) 020
[arXiv:hep-th/9906188].

\bibitem{Bianchi:2000hn}
M.~Bianchi, S.~Kovacs, G.~Rossi and Y.~S.~Stanev,
``Anomalous dimensions in N = 4 SYM theory at order g**4,''
Nucl.\ Phys.\ B {\bf 584} (2000) 216
[arXiv:hep-th/0003203].

\bibitem{Arutyunov:2002jg}
G.~Arutyunov and E.~Sokatchev,
``A note on the perturbative properties of BPS operators,''
Class.\ Quant.\ Grav.\  {\bf 20} (2003) L123
[arXiv:hep-th/0209103].

\bibitem{Bianchi:2003eg}
M.~Bianchi, G.~Rossi and Y.~S.~Stanev,
``Surprises from the resolution of operator mixing in N = 4 SYM,''
Nucl.\ Phys.\ B {\bf 685} (2004) 65
[arXiv:hep-th/0312228].

\bibitem{Dobrev:1985qv}
V.~K.~Dobrev and V.~B.~Petkova,
``All Positive Energy Unitary Irreducible Representations Of Extended Conformal
Supersymmetry,''
Phys.\ Lett.\ B {\bf 162} (1985) 127.\\
L.~Andrianopoli, S.~Ferrara, E.~Sokatchev and B.~Zupnik,
``Shortening of primary operators in N-extended SCFT(4) and
harmonic-superspace analyticity,''
Adv.\ Theor.\ Math.\ Phys.\  {\bf 3} (1999) 1149
[arXiv:hep-th/9912007].\\
P.~Heslop and P.~S.~Howe,
``On harmonic superspaces and superconformal fields in four dimensions,''
Class.\ Quant.\ Grav.\  {\bf 17} (2000) 3743
[arXiv:hep-th/0005135].

\bibitem{Ferrara:2000eb}
S.~Ferrara and E.~Sokatchev,
``Superconformal interpretation of BPS states in AdS geometries,''
Int.\ J.\ Theor.\ Phys.\  {\bf 40} (2001) 935
[arXiv:hep-th/0005151].

\bibitem{SemiShort}
G.~Arutyunov, S.~Frolov and A.~C.~Petkou,
``Operator product expansion of the lowest weight CPOs in N = 4  SYM(4) at
strong coupling,''
Nucl.\ Phys.\ B {\bf 586} (2000) 547
[Erratum-ibid.\ B {\bf 609} (2001) 539]
[arXiv:hep-th/0005182];\\
G.~Arutyunov, B.~Eden and E.~Sokatchev,
``On non-renormalization and OPE in superconformal field theories,''
Nucl.\ Phys.\ B {\bf 619} (2001) 359
[arXiv:hep-th/0105254];\\
B.~Eden and E.~Sokatchev,
``On the OPE of 1/2 BPS short operators in N = 4 SCFT(4),''
Nucl.\ Phys.\ B {\bf 618} (2001) 259
[arXiv:hep-th/0106249];\\
P.~J.~Heslop and P.~S.~Howe,
``A note on composite operators in N = 4 SYM,''
Phys.\ Lett.\ B {\bf 516} (2001) 367
[arXiv:hep-th/0106238];\\
S.~Penati and A.~Santambrogio,
``Superspace approach to anomalous dimensions in N = 4 SYM,''
Nucl.\ Phys.\ B {\bf 614} (2001) 367
[arXiv:hep-th/0107071];\\
F.~A.~Dolan and H.~Osborn,
``On short and semi-short representations for four dimensional superconformal
symmetry,''
Annals Phys.\ {\bf 307} (2003) 41
[arXiv:hep-th/0209056].

\bibitem{Park:1997bq}
J.~H.~Park,
``N = 1 superconformal symmetry in 4-dimensions,''
Int.\ J.\ Mod.\ Phys.\ A {\bf 13} (1998) 1743
[arXiv:hep-th/9703191].\\
H.~Osborn,
``N = 1 superconformal symmetry in four-dimensional quantum field theory,''
Annals Phys.\  {\bf 272} (1999) 243
[arXiv:hep-th/9808041].

\bibitem{Bianchi:2002rw}
M.~Bianchi, B.~Eden, G.~Rossi and Y.~S.~Stanev,
``On operator mixing in N = 4 SYM,''
Nucl.\ Phys.\ B {\bf 646} (2002) 69
[arXiv:hep-th/0205321].

\bibitem{Ryzhov:2001bp}
A.~V.~Ryzhov,
``Quarter BPS operators in N = 4 SYM,''
JHEP {\bf 0111} (2001) 046
[arXiv:hep-th/0109064].

\bibitem{Eden:1999gh}
B.~Eden, P.~S.~Howe and P.~C.~West,
``Nilpotent invariants in N = 4 SYM,''
Phys.\ Lett.\ B {\bf 463} (1999) 19
[arXiv:hep-th/9905085].

\bibitem{Penati:1999ba}
S.~Penati, A.~Santambrogio and D.~Zanon,
``Two-point functions of chiral operators in N = 4 SYM at order g**4,''
JHEP {\bf 9912} (1999) 006
[arXiv:hep-th/9910197].

\bibitem{Penati:2000zv}
S.~Penati, A.~Santambrogio and D.~Zanon,
``More on correlators and contact terms in N = 4 SYM at order g**4,''
Nucl.\ Phys.\ B {\bf 593} (2001) 651
[arXiv:hep-th/0005223].

\bibitem{HoweWest}
P.~S.~Howe and P.~C.~West,
``Superconformal invariants and extended supersymmetry,''
Phys.\ Lett.\ B {\bf 400} (1997) 307
[arXiv:hep-th/9611075].

\bibitem{Anselmi:1996mq}
D.~Anselmi, M.~T.~Grisaru and A.~Johansen,
``A Critical Behaviour of Anomalous Currents, Electric-Magnetic Universality
and CFT$_4$,''
Nucl.\ Phys.\ B {\bf 491} (1997) 221
[arXiv:hep-th/9601023].

\bibitem{Galperin:1984av}
A.~Galperin, E.~Ivanov, S.~Kalitsyn, V.~Ogievetsky and E.~Sokatchev,
``Unconstrained N=2 Matter, Yang-Mills And Supergravity Theories In Harmonic
Superspace,''
Class.\ Quant.\ Grav.\  {\bf 1} (1984) 469.

\bibitem{Mincer}K.~G.~Chetyrkin, A.~L.~Kataev and F.~V.~Tkachov,
``New Approach To Evaluation Of Multiloop Feynman Integrals: The Gegenbauer
Polynomial X Space Technique,''
Nucl.\ Phys.\ B {\bf 174} (1980) 345.\\
K.~G.~Chetyrkin and F.~V.~Tkachov,
``Integration By Parts: The Algorithm To Calculate Beta Functions In 4 Loops,''
Nucl.\ Phys.\ B {\bf 192} (1981) 159.\\
D.~I.~Kazakov,
``The Method Of Uniqueness, A New Powerful Technique For Multiloop
Calculations,''
Phys.\ Lett.\ B {\bf 133} (1983) 406.\\
S.~A.~Larin, F.~V.~Tkachov and J.~A.~M.~Vermaseren,
``The FORM version of MINCER,''
NIKHEF-H-91-18

\bibitem{Dubna} A.~A.~Vladimirov,
``Method For Computing Renormalization Group Functions In Dimensional
Renormalization Scheme,''
Theor.\ Math.\ Phys.\  {\bf 43} (1980) 417
[Teor.\ Mat.\ Fiz.\  {\bf 43} (1980) 210];\\
D.~I.~Kazakov, O.~V.~Tarasov and A.~A.~Vladimirov,
``Calculation Of Critical Exponents By Quantum Field Theory Methods,''
Sov.\ Phys.\ JETP {\bf 50} (1979) 521
[Zh.\ Eksp.\ Teor.\ Fiz.\  {\bf 77} (1979) 1035];\\
O.~V.~Tarasov and A.~A.~Vladimirov,
``Three Loop Calculations In Nonabelian Gauge Theories,''
JINR-E2-80-483.






\bibitem{yascom}
Y.~S.~Stanev, private communication.

\bibitem{Siegel:1980qs}
W.~Siegel,
``Inconsistency Of Supersymmetric Dimensional Regularization,''
Phys.\ Lett.\ B {\bf 94} (1980) 37.

\bibitem{Beisert:2004di}
N.~Beisert, M.~Bianchi, J.~F.~Morales and H.~Samtleben,
``Higher spin symmetry and N = 4 SYM,''
JHEP {\bf 0407} (2004) 058
[arXiv:hep-th/0405057].

\bibitem{Eden:2000mv}
B.~Eden, C.~Schubert and E.~Sokatchev,
``Three-loop four-point correlator in N = 4 SYM,''
Phys.\ Lett.\ B {\bf 482} (2000) 309
[arXiv:hep-th/0003096];\\
``Four-point functions of chiral primary operators in N = 4 SYM,''
[arXiv:hep-th/0010005].

\bibitem{harmonic}
A. Galperin, E. Ivanov, V. Ogievetsky, E. Sokatchev,
``Harmonic superspace", Cambridge University Press, Cambridge, 2001, 306 p.

\end{thebibliography}
\end{document}